\documentclass[
reprint,
superscriptaddress,
amsmath,amssymb,
aps,
prx,
]{revtex4-2}

\usepackage{graphicx}
\usepackage{dcolumn}
\usepackage{bm}
\usepackage{nicefrac}
\usepackage[colorlinks=true, allcolors=blue]{hyperref}
\begin{document}

\preprint{APS/123-QED}

\title{Transition metal ion ensembles in crystals as solid-state coherent spin-photon interfaces: The case of nickel in magnesium oxide}

\author{E. Poem}
\email{eilon.poem@weizmann.ac.il}
\affiliation{Department of Physics of Complex Systems, Weizmann Institute of Science, Rehovot 7610001, Israel}
\author{S. Gupta}
\affiliation{Prizker School of Molecular Engineering, The University of Chicago, Chicago, IL 60637, USA}
\author{I. Morris}
\author{K. Klink}
\affiliation{Department of Physics and Astronomy, Michigan State University, 567 Wilson Rd., East Lansing, MI 48824, USA}
\author{L. Singh}
\affiliation{Department of Physics of Complex Systems, Weizmann Institute of Science, Rehovot 7610001, Israel}
\author{T. Zhong}
\affiliation{Prizker School of Molecular Engineering, The University of Chicago, Chicago, IL 60637, USA}
\author{ S. S. Nicley}
\affiliation{ Department of Electrical and Computer Engineering, Michigan State University, 428S. Shaw Ln., Rm. 2120, East Lansing, MI, 48824, USA }
\affiliation{Coatings and Diamond Technologies Division, Center Midwest (CMW), Fraunhofer USA Inc., 1449 Engineering Research Ct., East Lansing, MI 48824, USA}
\author{J. N. Becker}
\affiliation{Department of Physics and Astronomy, Michigan State University, 567 Wilson Rd., East Lansing, MI 48824, USA}
\affiliation{Coatings and Diamond Technologies Division, Center Midwest (CMW), Fraunhofer USA Inc., 1449 Engineering Research Ct., East Lansing, MI 48824, USA}
\author{O. Firstenberg}
\affiliation{Department of Physics of Complex Systems, Weizmann Institute of Science, Rehovot 7610001, Israel}


\date{\today}

\begin{abstract}
We present general guidelines for finding solid-state systems that could serve as coherent electron spin -- photon interfaces even at relatively high temperatures, where phonons are abundant but cooling is easier, and show that transition metal ions in various crystals could comply with these guidelines. As an illustrative example, we focus on divalent nickel ions in magnesium oxide. We perform electron spin resonance spectroscopy and polarization-sensitive magneto-optical fluorescence spectroscopy of a dense ensemble of these ions and find that (i) the ground-state electron spin stays coherent at liquid-helium temperatures for several microseconds, and (ii) there exists energetically well-isolated excited states which can couple to two ground state spin sub-levels via optical transitions of orthogonal polarizations. The latter implies that fast, coherent optical control over the electron spin is possible. We then propose schemes for optical initialization and control of the ground-state electron spin using polarized optical pulses, as well as two schemes for implementing a noise-free, broadband quantum-optical memory at near-telecom wavelengths in this material system.
\end{abstract}

\maketitle


\section{Introduction}
Quantum communication and networks require coherent coupling between traveling qubits, carrying the quantum information, and stationary qubits, which can act as memories and processors, storing and manipulating the quantum information~\cite{Kimble2008TheInternet}. Photons at wavelengths in the telecommunication bands are arguably the best traveling qubits, as they have very low loss probability and can maintain their coherence over long distances in standard telecom fibers. Solid-state defect spins~\cite{ZhongREReview,Wolfowicz2021QubitDefects} are promising stationary qubits, as they are embedded in a miniaturizable platform. At room temperature, they can have coherence times as long as milliseconds~\cite{Balasubramanian2009UltralongDiamond} for electronic spins and tens of minutes~\cite{Saeedi2013minuteCoherenceSi} for nuclear spins. This only improves at low temperatures, where coherence times of up to seconds~\cite{BarGill2013oneSecondNV} and hours~\cite{Sellars2015SixHoursCoherence} have been achieved.  One of the outstanding challenges towards the realization of optical quantum networks is the coherent coupling of telecom photons and solid-state qubits. For defect spins, this would mean the coherent coupling of light and spin.

As light (at optical wavelengths, in the far field) directly affects only the orbit of the electron but not its spin~\cite{AtomPhotonBook.Ch1}, and even less so the nuclear spin, any spin-photon coupling has to be mediated by additional internal interactions within the quantum system. For electron spins, this would be the relativistic effect of spin-orbit (SO) coupling. For nuclear spins, one has to consider also the hyperfine coupling. In this work we discuss the case of electronic spins, where this leads to the two following requirements. 

First, to  coherently couple light at a certain frequency to an electron spin at a given temperature, that is, to coherently transfer quantum information between them, there should exist an excited state at that frequency (with respect to the ground state) in which the SO coupling rate is much faster than the total (homogeneous and inhomogeneous) decoherence rate of the ensemble at that temperature. 
This also includes cases where only a part of the ensemble is addressed, for example, by spectrally-selective optical pumping to a dark state~\cite{Corrielli2016RareEarthWGMemory}, or by using light with a narrower spectrum than that of the full ensemble~\cite{Hemmer2001NVRaman}, where the relevant decoherence rate would be that of the addressed sub-ensemble and not of the entire ensemble. Further, the duration of the optical field in the material should be long enough to allow for the SO interaction to act, but still shorter than the optical coherence time of the relevant ensemble. While some coherent effects have been previously seen in both ensembles~\cite{Rand1994fsFWMNV,Lenef1996NVFWM} and single spin centers~\cite{Awschalom2017FastControl} when using pulses shorter than the SO coupling time, these were due to coherent orbit-photon coupling, and not to coherent spin-photon coupling. 

For ensembles of spin defects, the typical total broadenings of the optical transitions are between 100~MHz to 100-GHz, at low temperatures~\cite{ZhongREReview,Weinzetl2019CoherentDiamond}. Thus, coherent spin-photon coupling using the entire ensemble thus requires \emph{SO coupling at the GHz-to-THz scale}. At higher temperatures, this may further increase. 

Second, at operating temperatures of few-kelvin and above, where the cooling power of existing cryostats dramatically increases, and their complexity and cost dramatically decrease~\cite{DilFridgeReview2022}, phonon modes of the defect and the host crystal are usually no longer frozen, as in mK temperatures~\cite{ZhongREReview}. Thus, the ground-state spin should couple to the electron orbit as little as possible, as such coupling would expose it to decoherence due to the interactions of the orbital degree of freedom with phonons. One known way of achieving this exists when the quantum information is encoded on nuclear spins. In that case, one can apply a strong magnetic field and shift the energy of neighboring electronic-spin states enough to quench two-phonon scattering processes even at few-kelvin temperatures~\cite{Bottger2009MagneticFieldT2,Thiel2010MagneticFieldT2,Rancic2018secondT2}. 

An alternative solution, which does not require high magnetic fields, nor sub-kelvin temperatures, and applies to electronic spin qubits, is to use an electronic system with a \emph{ground state orbital singlet}, and therefore zero effective orbital angular momentum and zero first-order SO coupling in the ground state. \color{black}

The tension between these two requirements can be illustrated with a few example cases. 

The first case is that of ensembles of NV$^-$ centers in diamond~\cite{Doherty2013TheDiamond}. While these ensembles comply with the second requirement ($^3$A$_2$ ground state manifold), they do not comply with the first ( $\sim$30~GHz inhomogeneous broadening and only $\sim$3~GHz SO coupling within the $^3$E excited state manifold). Therefore, they have excellent spin properties~\cite{Balasubramanian2009UltralongDiamond,BarGill2013oneSecondNV} but only incoherent spin-photon coupling (coherent spin-photon coupling was achieved only at low temperatures, for single NV$^-$ centers~\cite{Lukin2010Entanglement,Awschalom2013Control,Wang2014NVOptCont,Hensen2015Loophole,Awschalom2017Holonomic,Awschalom2017FastControl} or for very small sub-ensembles~\cite{Hemmer2001NVRaman}\color{black}). 

The second case includes both negatively-charged group-IV-vacancy centers in diamond~\cite{Chen2020GroupIVcentersReview} and commonly-used rare-earth ion ensembles~\cite{Bottger2009MagneticFieldT2,Thiel2010MagneticFieldT2,Thiel2011Rare-earth-dopedProcessing,Rancic2018secondT2,ZhongREReview}, like Pr$^{3+}$, Eu$^{3+}$, Yb$^{3+}$ or Er$^{3+}$. These systems feature orthogonal challenges to the NV$^-$ center, as they have strong SO in the excited state, but an orbital multiplet ground state manifold. Therefore, while they may support high-temperature coherent spin-photon coupling, in the absence of very high magnetic fields their ground state spin quickly decoheres above liquid helium temperatures. For example, the electron spin coherence time of Yb$^{3+}$ in $\mathrm{Y_2 Si O_5}$ at 9~K is 2 $\mathrm{\mu s}$, limited by fast two-phonon spin relaxation~\cite{ybysopgoldner}. \color{black}

The third example case is that of the three known zero orbital angular momentum (S-state) rare-earth ions, Gd$^{3+}$~\cite{DEBIASI20041207,Hughes1974Gd3CaO}, Eu$^{2+}$~\cite{Ebendorff1999Eu2Tb4ESR,PAN200678}, and Tb$^{4+}$~\cite{Ebendorff1999Eu2Tb4ESR,VERMA20101146}, and the S-state actinide ion Cm$^{3+}$~\cite{ILLEMASSENE2000CmSpectroscopy}. These ions comply with both requirements and indeed exhibit very narrow spin distributions and optical line widths, some of them even up to room temperature. Unfortunately, the first three have optical transitions only in the ultra-violet, and the last one, while having transitions in the visible, is radioactive. This makes them less suitable for optical communications purposes.

In contrast to the examples above, transition metal ions, even in high symmetry lattice sites of cubic crystals~\cite{Shang2022CalcTMionsQIP}, can comply with both requirements while having optical transitions at infrared or even telecom wavelengths. First, due to their relatively large atomic number, the SO interaction is usually on the order of a few THz (similarly to the case of rare earth ions). Second, due to the effect of the crystal field, which can be much stronger than in rare earth ions, there are multiple cases with ground state orbital singlets (similar to the NV$^-$). This happens whenever the highest set of degenerate single-electron orbitals is half-filled. For weak crystal fields, forming high-spin configurations, the only relevant configuration is $d^5$ (in analogy to the $f^7$ configuration of the S-state rare earth ions). For stronger crystal fields, however, where low-spin configurations form, there are more options. For cubic sites, these include $d^8$ and $d^3$ ions in octahedral sites, and $d^2$ and $d^7$ ions in tetrahedral sites. Of these, the configurations with an even number of electrons have an $S=1$ ground state, while those with an odd number have an $S=\nicefrac{3}{2}$ ground state.
While some transition metal ion systems were investigated in the context of quantum information processing~\cite{Bosma2018IdentificationCarbide,Gilardoni2020MoSiC,Wolfowicz2020vanadium4,Astner2022V4}, the only such system having an orbital singlet ground state that has been investigated in this context, to the best of our knowledge, is Cr$^{4+}$ in GaN and SiC~\cite{Koehl2017ResonantGaN,Diler2020CoherentCarbide}, having a $d^2$ configuration in a tetrahedral site. It indeed displays infrared emission (around 1090 nm) and a coherent ground state spin up to at least 15~K. This spin could potentially be coherently controlled optically, though, to the best of our knowledge, this has not been demonstrated yet.

Here we focus on another such example: divalent nickel (Ni$^{2+}$) substituting for a magnesium ion in magnesium oxide (MgO). It has a $d^8$ configuration, and due to the octahedral geometry of its site, its ground state is an orbital singlet and spin-triplet. This system was very thoroughly studied in the past few decades, both for fundamental characterization~\cite{Low1958NiMgO,Orton1960NiMgOESR,Walsh1961StrainESR,Orton1961Ni61MgOESR,Pappalardo1961OpticalI,Minomura1961pressure,Lewis1967phononESR,Lewis1967T1,Ralph1968Near-infraredMgO,Smith1969EPRMgO,Ralph1970FluorescenceMgO,Manson1971One-phononMgO,Sangster1970PhononTheory,Bird1972NiMgOMCD,Wong1973NiMgOMCD,Moreau19741T2gJT,Manson19761T2gSplitting,Thorp1981ESRNiMgO,Payne1990NiMgOexcited,Campochairo1991MgONiTwoPhot,Mironova1996X-raySolutions,Mironova-Ulmane2013CrystalMgO} and for applications, mostly as a potential gain medium for tunable and pulsed infrared lasers~\cite{Moodrian1979NiMgF2Laser,Iverson1980NiMgOLaserParameters,Moncorge1988ESA}. The spin triplet nature of the ground state was confirmed by electron spin resonance (ESR) studies already over 60 years ago~\cite{Low1958NiMgO,Orton1960NiMgOESR,Walsh1961StrainESR}. The spin-lattice relaxation time (T$_1$) was measured to be as long as 1~ms at 3.5~K and 17~$\mu$s at 35~K for a magnetic-field-induced ground-state splitting of 9.2~GHz. Even longer times may be measured for smaller splittings, especially at the lower part of the temperature range (20~K and below), where the dominant process is single-phonon scattering, the rate of which scales quadratically with the spin splitting~\cite{Lewis1967phononESR,Lewis1967T1}. Optical studies revealed rich emission and absorption spectra~\cite{Pappalardo1961OpticalI,Ralph1968Near-infraredMgO,Ralph1970FluorescenceMgO,Moncorge1988ESA,Payne1990NiMgOexcited,Campochairo1991MgONiTwoPhot}. In particular, the lowest emission energy zero-phonon lines (ZPLs) are   at 1220~nm and 1250~nm, where the loss rate in a commercial optical fiber is $\sim$0.4~dB/km, not much different than the $\sim$0.3~dB/km loss rate in the O-band (1310~nm).   Remarkably, these lines remain well separated up to temperatures as high as 150~K. The optical lifetime of these lines is very long, about 3.6~ms~\cite{Iverson1980NiMgOLaserParameters,Moncorge1988ESA}, up to temperatures of $\sim$100~K. While this indicates a weak transition dipole moment, predominantly magnetic, due to the perfect solubility of NiO in MgO~\cite{Mironova1996X-raySolutions}, optically dense ion ensembles could be readily made, compensating for the weak optical response of individual ions and enabling a strong collective response. Furthermore, as only 5\% of the atoms in MgO made with natural isotope abundances have non-zero nuclear spin (due to $^{25}$Mg), the spin dephasing rates due to nuclear spin-bath noise should be low and were theoretically predicted to be below 1~kHz~\cite{Cheng2017DivalentApplication,Awschalom2022MaterialSurvey}. Importantly, like diamond~\cite{Balasubramanian2009UltralongDiamond}, this material can be made nuclear-spin-free by using isotopically purified precursors\color{red}~\cite{Catanzaro1966PureMgO}\color{black}.

These compelling features lead us to re-examine this material system for use as a coherent spin-photon interface at above-liquid helium temperatures. We experimentally investigate both the ground-state spin decoherence mechanisms and the spin structure of the excited state, and show that they are compatible with THz bandwidth coherent optical spin control even at temperatures exceeding that of liquid helium,   estimated to go up to a few tens of K.  

The rest of the manuscript is organized as follows. In Sect.~\ref{sect:levels}, we describe the level structure of Ni$^{2+}$ in MgO, both of the ground state and of the excited states, where for the latter we focus on the difference between the case of a weakly-perturbed SO coupling and that of quenched SO coupling due to a strong dynamic Jahn-Teller (DJT) distortion of the excited state orbitals~\cite{Ham1965DJT}. In Sect.~\ref{sect:spin}, using ESR and temperature-dependent spin-echo (SE) measurements, we show that for a high-density ensemble, the main decoherence mechanism is dipolar interaction between the ensemble spins. For a few GHz of ground state splitting (induced by an external magnetic field), this interaction is already saturated at liquid-helium temperatures, leading to a decoherence time of 3~$\mu$s. 
In Sect.~\ref{sect:opt_spect}, we use polarization-sensitive magneto-optical fluorescence spectroscopy on the two ZPLs and show that the excited-state spin structure is indeed determined mostly by the SO interaction since the DJT distortion of the excited-state orbitals is weak. Following these findings, in Sect.~\ref{sect:proposal}, we propose protocols for optical spin-state preparation, measurement, and manipulation, as well as for noise-free optical quantum memories. Finally in Sect.~\ref{sect:conclusion}, we summarize our results and outline possible directions for future research.

\section{level structure}\label{sect:levels}
\subsection{Ground state}
\begin{figure*}[tbh]
\centering
\includegraphics[width=0.99\textwidth]{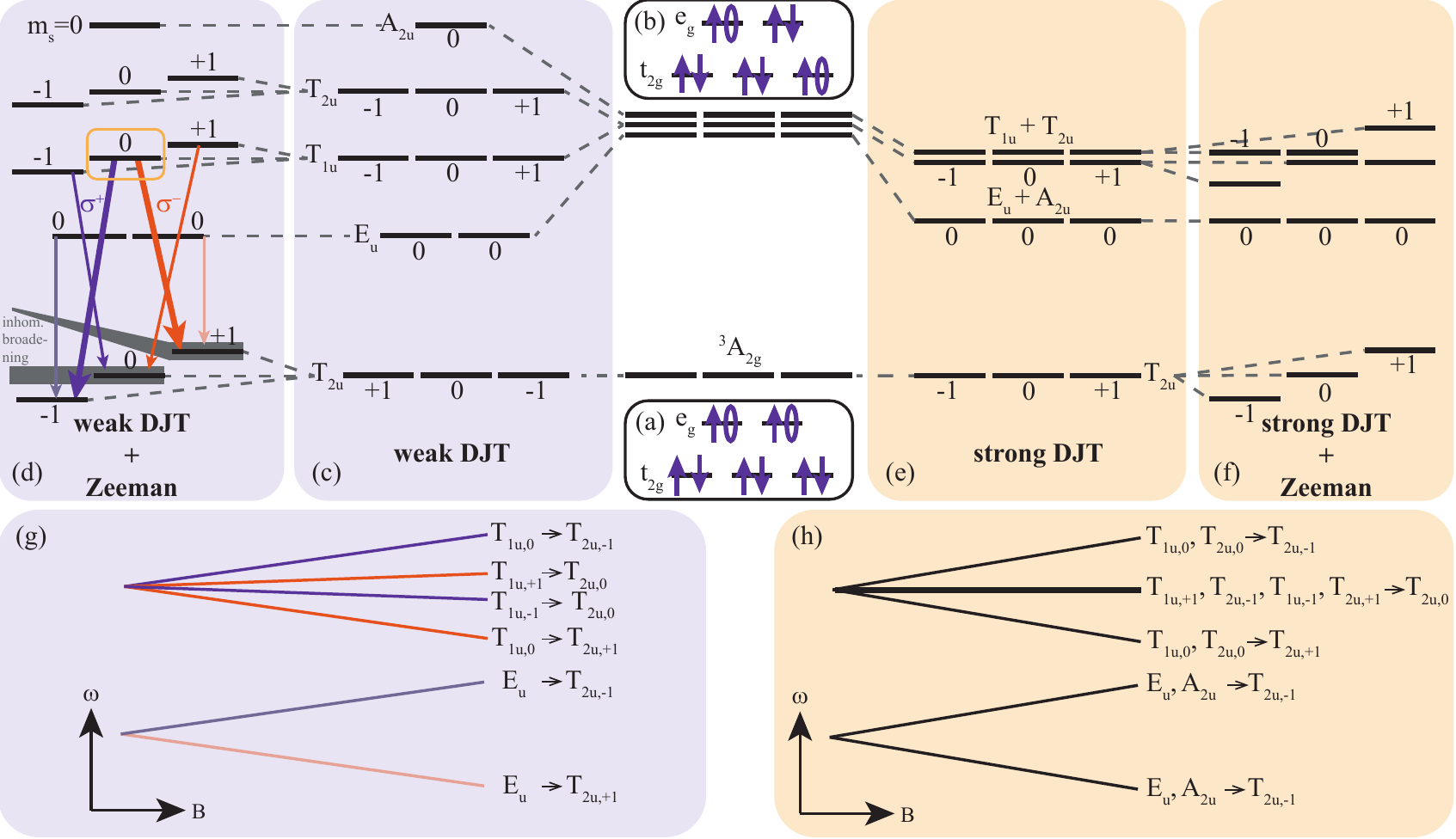}
\caption{\label{fig:levels} Energy level structure of Ni$^{2+}$:MgO. (a) Ground-state electronic configuration. The arrows represent electrons, and the ellipses represent holes. (b) Electronic configuration of the first excited state. (c) [(e)] Ground and excited state level structures resulting from SO interaction in the presence of a weak [strong] DJT. (d) [(f)] Splitting of the ground and excited state spin sub-levels in a magnetic field for weak [strong] DJT. The grey shaded regions in the weak DJT ground state represent random-strain-induced inhomogeneous broadenings and their field dependencies. A polarized $\Lambda$-system (orange rectangle) forms for weak DJT. Other relevant optical transitions are marked, where purple (orange) arrows represent $\sigma^+$ ($\sigma^-$) polarized transitions, and faded-colored lines arrows represent partially-polarized transitions. (g) [(h)] Splitting of the optical emission lines under a magnetic field (field increases to the right) in Faraday configuration, for the case of weak [strong] DJT. The same color coding as in (d) is used. Black lines denote unpolarized transitions. $\pi$-polarized transitions are not shown.}
\end{figure*}

The ground state of a Ni$^{2+}$ ion in an octahedral site of a cubic lattice (O$_h$ symmetry group) contains two electrons (or, equivalently, two electron-holes) occupying two degenerate $e_g$ single-electron orbitals (formed by the $d_{x^2-y^2}$ and $d_{z^2}$ $d$-orbitals), as shown in Fig.~\ref{fig:levels}(a). As this is a half-filled shell, there is only one many-electron orbital. The total spin of the two electrons can be either 0 or 1, but the spin-1 states have a lower energy. The ground state is therefore an orbital-singlet spin-triplet, $^3$A$_{2g}$, the spin-orbit representation of which is $T_{2u}$. The effective Hamiltonian for this manifold is 
\begin{equation}\label{eq:gs_H}   
H^{\mathrm{g}}=\mu_{\mathrm{B}} g^{\mathrm{g}}_s \mathbf{B\cdot S}+\mathbf{S\cdot q(\epsilon)\cdot S}+\mu_{\mathrm{B}} \mathbf{B\cdot \delta g_S(\epsilon)\cdot S},
\end{equation}
where $\mathbf{S}$ is the vector of spin-1 operators, $\mathbf{B}$  is the magnetic field vector, $\mathbf{q}(\epsilon)$ is the strain-induced magnetic quadrupole moment, $\mathbf{\delta g_S}(\epsilon)$ is the strain-induced g-tensor~\cite{Rosenberg1967strain}, and $\epsilon$ is the strain tensor. Here $\mu_\mathrm{B}$ is the Bohr magneton, and $g^\mathrm{g}_s$ is the ground state g-factor. The three spin states are degenerate at zero fields and strains. A constant magnetic field along the z-direction splits the states via the Zeeman interaction [first term in Eq.~(\ref{eq:gs_H})]. Due to the second term in Eq.~(\ref{eq:gs_H}), local random strains (of a cubic lattice) can shift the $T_{2u,0}$ state with respect to the $T_{2u,\pm1}$ states to first order in the ratios of the strain energies and the Zeeman energy. This inhomogeneously broadens the $T_{2u,1}\leftrightarrow T_{2u,0}$ and the $T_{2u,0}\leftrightarrow T_{2u,-1}$ spin transitions. The same term can also split the $T_{2u,\pm1}$ states, however only to second order in strain-to-Zeeman energy ratio~\cite{Rosenberg1967strain}. This is because the relevant strain terms are off-diagonal in the magnetic-field Hamiltonian, such that their effect is quenched as the magnetic field becomes large, leading to a reduced broadening of the $T_{2u,1}\leftrightarrow T_{2u,-1}$ spin transition. The third term in Eq.~(\ref{eq:gs_H}) cannot shift the $T_{2u,0}$ state with respect to the $T_{2u,\pm1}$ states, but can split the $T_{2u,\pm1}$ states to first order. However, in MgO, for an applied magnetic field on the order of 100~mT (few GHz $T_{2u,1}\leftrightarrow T_{2u,0}$ Zeeman splitting), this splitting is about two orders of magnitude smaller than the first-order shifts induced by the second term~\cite{Walsh1961StrainESR,Rosenberg1967strain,Mattuck1960strain,Zheng1989strain,Ma1998strain_calc}, keeping the broadening of the $T_{2u,1}\leftrightarrow T_{2u,-1}$ spin transition smaller than that of the $T_{2u,0}\leftrightarrow$$T_{2u,\pm1}$ transitions. The level splitting and broadenings are schematically presented in Fig.~\ref{fig:levels}(d). In Sect.~\ref{sect:spin} below, we present ESR measurements of these inhomogeneous broadenings, as well as SE measurements of the homogeneous decoherence time at different temperatures.  

\subsection{Excited state}
The first excited state of Ni$^{2+}$ in MgO is composed of one electron-hole in one of two $e_g$ single-electron orbitals, and one hole in one of three  $t_{2g}$ single-electron orbitals (formed by the $d_{yz}$, $d_{xz}$, and $d_{xy}$ $d$-orbitals), as shown in Fig.~\ref{fig:levels}(b). Therefore, there are six possible two-electron orbitals. The cubic symmetry splits these six states into two orbital triplets, T$_{1g}$ and T$_{2g}$, where the latter has a lower energy~\cite{Pappalardo1961OpticalI}. The spin state of the two electrons in the lowest excited-state manifold is again a spin-1 triplet. Thus, the lowest excited-state manifold is $^3$T$_{2g}$, which contains 9 states in total. 

These states are coupled and split by the spin-orbit interaction. Its magnitude depends on the strength of the dynamic Jahn-Teller (DJT) coupling between the electronic orbitals and lattice vibrations, as the latter may affect the shapes of the orbitals and thus their effective angular momentum~\cite{Ham1965DJT}.
The effective Hamiltonian for the $^3$T$_{2g}$ manifold, in the presence of a magnetic field, is
\begin{equation}\label{eq:es_H}
H^\mathrm{e}(\kappa)=H_\mathrm{so}^\mathrm{e}(\kappa)+H_B^\mathrm{e}(\kappa),
\end{equation}
where $\kappa=3E_\mathrm{JT}/\hbar\omega_\mathrm{ph}$ is the relative strength of the DJT coupling. It is proportional to the ratio between the electron-phonon coupling energy $E_\mathrm{JT}$ and the energy of the lowest phonon mode $\hbar\omega_\mathrm{ph}$. 

The zero-field Hamiltonian is~\cite{Ham1965DJT}
\begin{widetext}
\begin{equation}\label{eq:es_SO_DJT}
H_\mathrm{so}^\mathrm{e}(\kappa)=e^{-\kappa/2}\zeta\mathbf{L\cdot S}+[\mu e^{-\kappa/2}+K_1(\kappa)](\mathbf{L\cdot S})^2+[\rho+\mu(1-e^{-\kappa/2})+K_2(\kappa)]A,
\end{equation}
\end{widetext}
where $\mathbf{L}$ ($\mathbf{S}$) is the orbital (spin) angular momentum vector operator, and $A=L_x^2S_x^2+L_y^2S_y^2+L_z^2S_z^2$ is a second-order cubic-symmetry spin-orbit term. Here, both $\mathbf{L}$ and $\mathbf{S}$ are spin-1 operators. The quadratic SO terms (the last two terms) arise from second-order perturbation theory applied to the full Hamiltonian~\cite{Ham1965DJT}. The energies $K_1$ and $K_2$ are given by, $K_1=(g_L^2\zeta^2/\hbar\omega_\mathrm{ph})e^{-\kappa}G(\kappa/2)$, and $K_2=(g_L^2\zeta^2/\hbar\omega_\mathrm{ph})e^{-\kappa}[G(\kappa)-G(\kappa/2)]$, where $G(x)=\int_0^xdt(e^t-1)/t$~\cite{Ham1965DJT,Kaufmann1973DJT}. The magnetic-field dependent Hamiltonian is
\begin{equation}\label{eq:es_H_B}
H_B^\mathrm{e}=\mu_\mathrm{B}(g_L e^{-\kappa/2}\mathbf{L}+g^\mathrm{e}_s\mathbf{S})\cdot\mathbf{B},
\end{equation}
where $g_s^\mathrm{e}$ is the excited state g-factor, and $g_L$ its orbital gyro-magnetic ratio. Here we neglected the static strain shifts.\\

For weak DJT distortion, that is, for $\kappa<1$, the ``static lattice" SO structure survives, and the nine-fold degenerate state-space splits into four distinct energy levels (see table~\ref{tab:weak_sr} in Appendix~\ref{app:pol_sel_rules}): \color{black}
a doublet ($E_u$), two triplets ($T_{1u}$ and $T_{2u}$), and a singlet ($A_{2u}$), as shown in Fig.~\ref{fig:levels}(c). Out of these, the $T_{1u}$ triplet contains one state, $T_{1u,0}=\nicefrac{1}{\sqrt{2}}(|$T$_{2g,1}\rangle|1\rangle_s-|$T$_{2g,-1}\rangle|\mathrm{-}1\rangle_s)$ (marked in the figure) which mixes the $|\pm1\rangle_s$ spin states, each coupled to a different orbital. Note that the total angular momentum components in this state are $\pm2$, which reverses the polarization selection rules of the transitions from it to the $T_{2u,\pm1}$ ground states with respect to those expected from a zero-total-angular-momentum state. Nevertheless, these three states and the transitions between them manifest a polarized $\Lambda$ system [Fig.~\ref{fig:levels}(d)]. Such a level system enables the control of the $T_{2u,\pm1}$ ground-state two-level system using polarized light~\cite{Kodriano2012CompletePulse,Simon2014NVMemProp,Wang2014NVOptCont,Treutlein2022PolMem}.

For the opposite case of a strong DJT distortion, where $\kappa\gg1$, only the $A$ term in the Hamiltonian of Eq.~(\ref{eq:es_SO_DJT}) survives, and there are only two energy levels (see table~\ref{tab:strong_sr} in Appendix~\ref{app:pol_sel_rules}): a triplet (composed of $E_u$ and $A_{2u}$) and a sextuplet (composed of $T_{1u}$ and $T_{2u}$), as shown in Fig.~\ref{fig:levels}(e). In this case, due to destructive interference between different states of the same level, no $\Lambda$-system can form~\cite{Poem2015BroadbandDiamond}, and coherent optical spin manipulation is prohibited [Fig.~\ref{fig:levels}(f)].
It is therefore crucial to distinguish between the weak and strong DJT regimes. 

If the absorption spectrum would feature four distinct narrow lines, as is the case, for example, for Ni$^{2+}$ in forsterite~\cite{Walker1994NiForsterite}, the favorable weak DJT case would be clearly identified. However, only two of the observed absorption lines are narrow, while the rest are broad~\cite{Pappalardo1961OpticalI}. This could be for one of two reasons: (i) the broad lines include the two remaining ``static lattice" lines, but mixed with high-energy vibrations (which are not included in the above model), or (ii) the broad lines are purely vibrational lines, and the two narrow lines are the result of a strong DJT distortion of the electronic levels. 

In the literature, the common interpretation is that of weak DJT, and some studies assign energies to the upper two electronic transitions~\cite{Pappalardo1961OpticalI,Moncorge1988ESA}. However, so far, this interpretation has not been validated other than via a theoretical analysis of the possible vibrational modes of Ni$^{2+}$ in MgO~\cite{Sangster1970PhononTheory,Manson1971One-phononMgO}, and recent theoretical studies have questioned it~\cite{Mironova-Ulmane2013CrystalMgO}. In Appendix~\ref{app:DJTmodel}, we show that both the strong and weak DJT cases can fit the observed spectra. Furthermore, cases of other transition metal spin-1 systems where only two of the four expected narrow absorption lines were observed, namely V$^{3+}$ in GaAs, GaP, InP~\cite{Kaufmann1982VanGaPGaAs,Lambert1983VanInP,Armelles1984Uniaxial-stressArsenide,Armelles1984MagnetoSpectV3GaAs,Ulrici1985,Ulrici1987OpticalPhosphide,Hennel1987OpticalArsenide,Gorger1988IdentificationGaAs,Chen2000InvestigationInP,Bchetnia2003AOMVPE}, and ZnO~\cite{Heitz1991ZeemanZnO}, have been reported, and the strong DJT distortion case was shown to be valid for these systems. This was achieved by analyzing the magnetic-field dependence of the absorption spectrum~\cite{Armelles1984MagnetoSpectV3GaAs,Heitz1991ZeemanZnO}. For the case of Ni$^{2+}$ in MgO, while magnetic circular dichroism has been probed in the past~\cite{Bird1972NiMgOMCD,Wong1973NiMgOMCD}, no conclusion regarding the excited state spin structure was drawn. 

Figures \ref{fig:levels}(g) and \ref{fig:levels}(h) present the magnetic splitting of the optical transitions for the cases of weak and strong DJT, respectively (see Appendix~\ref{app:pol_sel_rules}). It is clearly seen that both the number of spectral components and their polarizations differ between the two cases, allowing for a clear distinction between them.
In Sect.~\ref{sect:opt_spect} below, we present polarized optical magneto-fluorescence spectroscopy measurements for the two lowest-energy ZPLs of Ni$^{2+}$ in MgO, which unambiguously support the case of a weak DJT distortion,   with $\kappa\approx0.13$ (see Appendix~\ref{app:DJTmodel}).  

\section{Ground state spin coherence}\label{sect:spin}
For the experiment, we use a 5$\times$5$\times$1~mm single-crystal MgO sample, cut along the (001) planes and optically-polished on the two large facets, grown by Goodfellow Inc. It was intentionally doped with 450~ppm (2.4$\times10^{19}$~cm$^{-3}$) of nickel (nominally) and had a nominal concentration of 10~ppm of unintentional dopants. The sample was investigated as grown, with no further processing.
\begin{figure}[tb]
\centering
\includegraphics[width=1\columnwidth]{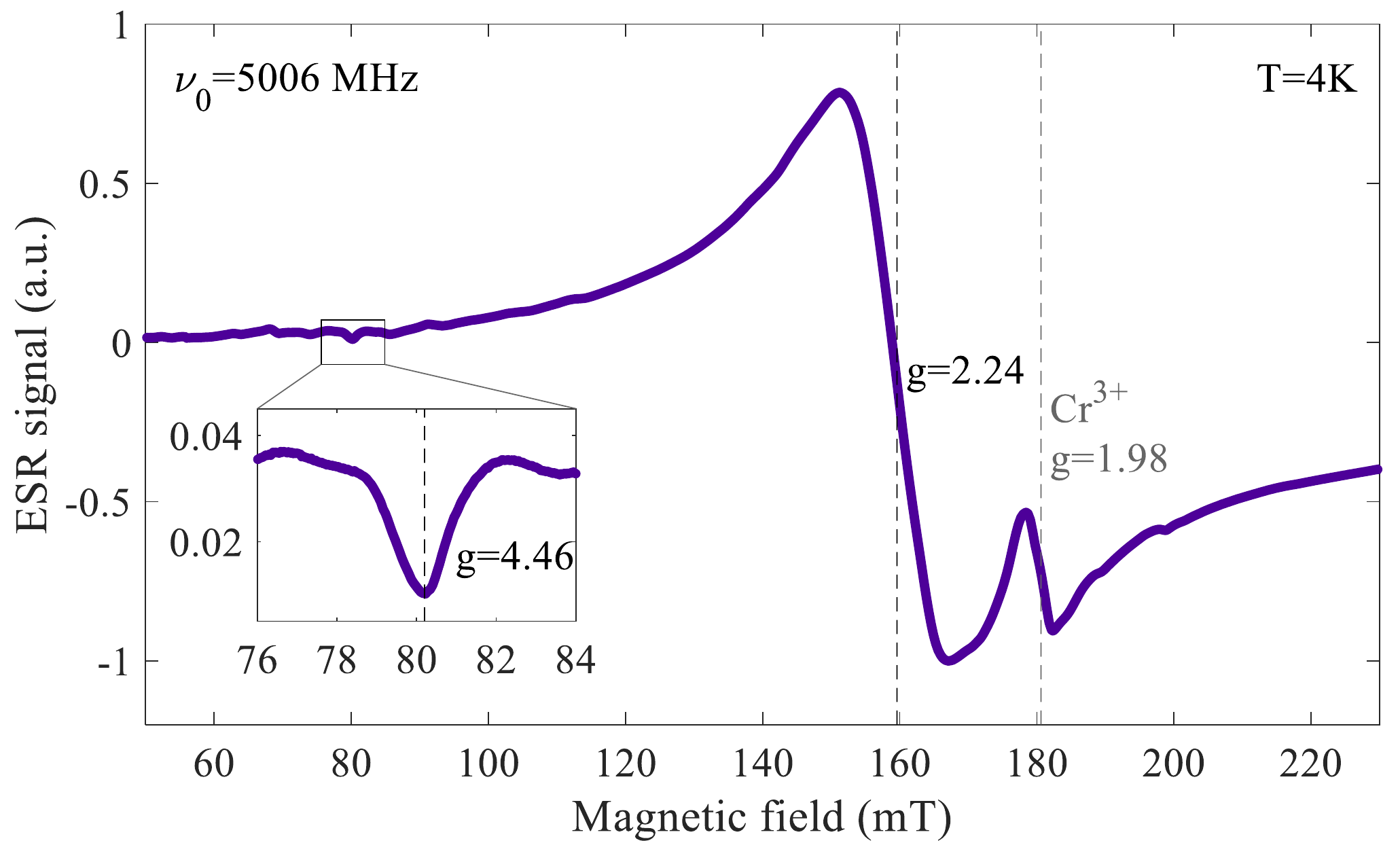}
\caption{\label{fig:ESR} Electron spin resonance (ESR) spectrum of the ground-state spin of Ni$^{2+}$:MgO. The magnetic field was calibrated using the Cr$^{3+}$ line at $g=1.98$. The inset magnifies the spectrum around the $\Delta m_s=2$ transition.}
\end{figure}

For ESR and SE measurement, the sample was placed in a 5.006~GHz aluminum microwave loop-gap cavity with a line-width of 1~MHz. The cavity was mounted on the mixing chamber of a Bluefors LD250 dilution refrigerator reaching a base temperature of 8~mK. A three-axis vector magnet (AMI Model 430) was used to apply a magnetic field. A cryogenic amplifier (Low noise factory LNF-LNC0.3\_14a) was used to pre-amplify the spin echo signal, limiting the maximum microwave power to $<1$~mW and the maximum temperature to 4~K.

The ESR spectrum measured at 4~K is presented in Fig.~\ref{fig:ESR} with the magnetic field approximately aligned along the [100] axis. The relatively narrow feature around 180~mT ($g=1.98$) is a well-known transition of Cr$^{3+}$~\cite{Hartman1970Cr3MgOESR}. We verified the presence of Cr$^{3+}$ in our sample also using fluorescence spectroscopy (see Sect.~\ref{sect:opt_spect} below). We used the known g-factor of the Cr$^{3+}$ transition to calibrate the magnetic field. 
The dominant broad feature around 160~mT ($g=2.24$) is related to Ni$^{2+}$. The measured g-factor matches the known value of 2.214~\cite{Orton1960NiMgOESR} to the precision of our magnetic field calibration. Its width (peak-to-peak) is 13~mT, corresponding to 400~MHz line width. This large broadening is probably due to random strain introduced by the high concentration of dopants, limiting the inhomogeneous coherence time of the spin ensemble to $\sim1$~ns. Similar widths have been measured previously, and it is also known that high-temperature annealing reduces the width by about a factor of two~\cite{Smith1969EPRMgO}. We did not see the `inverse line' previously observed in the center of the Ni$^{2+}$ line~\cite{Orton1960NiMgOESR,Smith1969EPRMgO}. As this line was attributed to a homogeneous, resonant cross-relaxation process~\cite{Smith1969EPRMgO}, it could be that this process was quenched due to the low temperature in our experiment. We also could not observe the `double-quantum' line, due to a two-photon transition between the $T_{2u,-1}$ and $T_{2u,1}$ states~\cite{Orton1960NiMgOESR,Lewis1967phononESR,Smith1969EPRMgO}, most probably due to our microwave-power limitation. 

In addition to the strong Ni$^{2+}$ and Cr$^{3+}$ lines, the measured spectrum exhibits a small feature very close to half the magnetic field of the main Ni$^{2+}$ resonance, as shown in the inset of Fig.~\ref{fig:ESR}. We attribute this feature, which has the form of a Fano resonance (`bound state in a continuum'~\cite{AtomPhotonBook.Ch1}), to the forbidden, $\Delta m_s=2$ single-photon transition between the $T_{2u,-1}$ and $T_{2u,1}$ states of the Ni$^{2+}$ ground state. The transition becomes partially allowed due to strain~\cite{Rosenberg1967strain}. The Fano shape, also seen in previous works~\cite{Lewis1967phononESR}, is most probably due to interference with the wide background coming from the $\Delta m_s=1$ transition. The width of this line (FWHM) is 1.5~mT, corresponding to about 90~MHz and $\sim5$~ns inhomogeneous coherence time. This narrow width (relative to the main transition) results from the lower strain sensitivity of the energy gap between the $T_{2u,\pm1}$ levels~\cite{Rosenberg1967strain}. A similar ratio was also measured for the double-quantum transition~\cite{Orton1960NiMgOESR,Lewis1967phononESR,Smith1969EPRMgO}.  

\begin{figure}[tb]
\centering
\includegraphics[width=1\columnwidth]{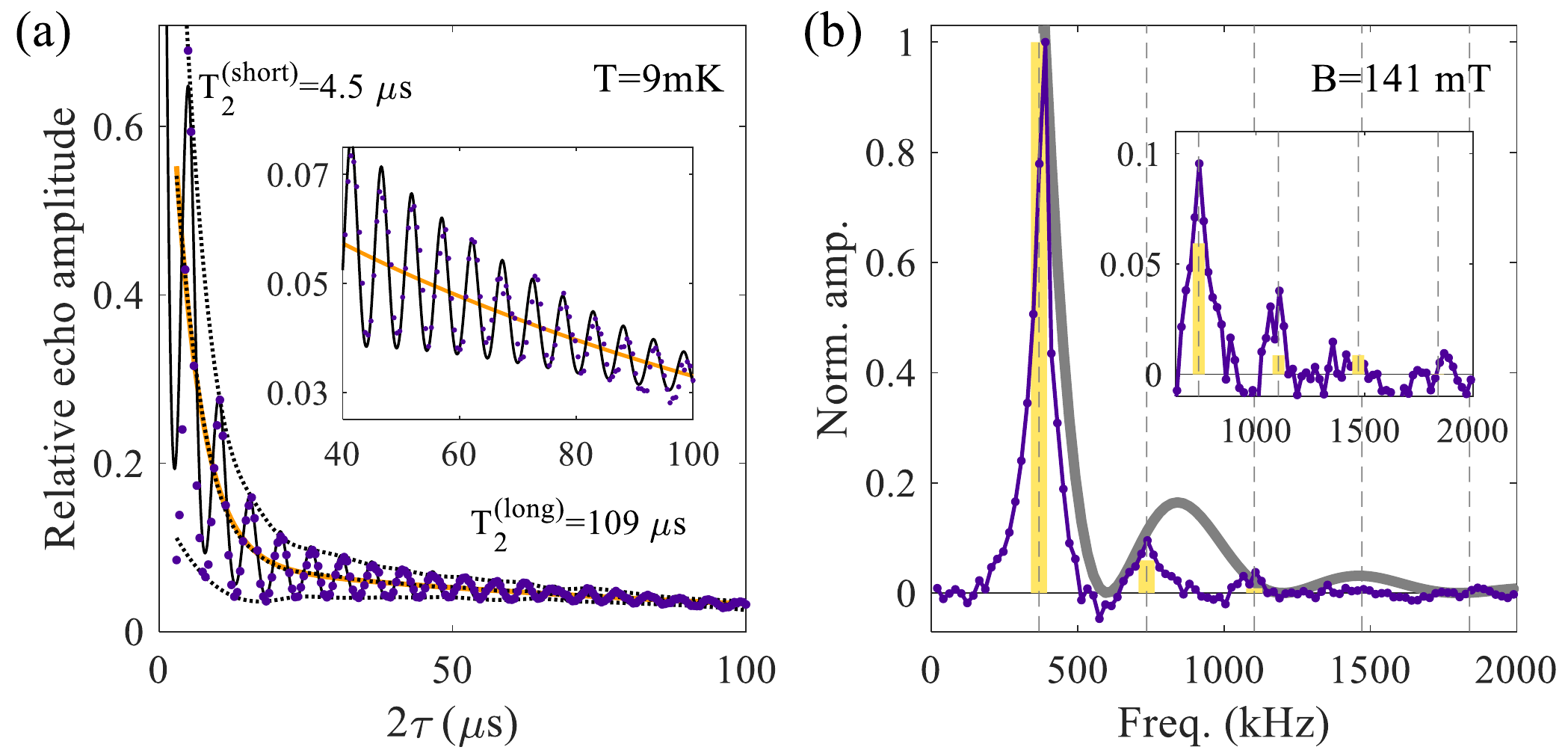}
\caption{\label{fig:echo} Spin echo at 9~mK and 141~mT. (a) Amplitude versus the time between the two driving pulses. The dotted lines show the oscillation envelopes and their mean. The orange line is a fit of the mean of the oscillation envelopes to a bi-exponential decay curve. The solid black line is the model fit (see text and Appendix~\ref{app:ESEEM}). The inset zooms-in on the long-time range. (b) The frequency content of the measured echo decay. A wide background peaked at zero frequency was subtracted. The yellow bars are the relative oscillation amplitudes predicted by the ESEEM model (see text and Appendix~\ref{app:ESEEM}) multiplied by the instrumental spectral response (gray line). The inset zooms-in on the high-frequency range. \color{black}}
\end{figure}

Next, we set the magnetic field to 141~mT, at the edge of the distribution, and measure the spin echo following excitation with two 500~ns-long pulses. The measured echo amplitude versus the time between the pulses, taken at 9~mK, is presented in Fig.~\ref{fig:echo}(a). Most strikingly, we observe pronounced oscillations. The Fourier transform of this pattern is presented in Fig.~\ref{fig:echo}(b). The main frequency component is at 385$\pm$10~kHz, fitting the predicted 366~kHz of nuclear Zeeman splitting of $^{25}$Mg at the applied field rather well. Additional components at the second and third harmonics of this frequency are also visible. The oscillations can therefore be explained as an electron spin-echo envelope modulation (ESEEM)~\cite{Mims1965ESEEM,Probst2020ESEEM} caused by the coupling of the Ni$^{2+}$ electronic spin to the nuclear spins of neighboring $^{25}$Mg atoms. The modulation frequencies exactly match multiples of the nuclear Zeeman splitting [marked by vertical dashed lines in Fig.~\ref{fig:echo}(b)] and are not affected by any hyper-fine coupling terms despite the strong modulation depth, in principle necessitating strong hyper-fine coupling. This can be explained by the zero spin component of the excited electronic state, limiting hyper-fine coupling to the ground state electron spin, which is fully occupied at the experiment temperature. However, as the nuclear spin-state is still fully mixed at the experiment temperature, only transitions with the same ground-state nuclear spin and different excited state nuclear spin would interfere, and thus hyper-fine coupling does not show up in the modulation frequency. Using the model presented in Refs.~\cite{Mims1965ESEEM,Probst2020ESEEM}, adapted to the case of initial (final) electron spin component of $-1$ (0) and a nuclear spin of 5/2 (see Appendix~\ref{app:ESEEM}), we calculate the expected modulation frequencies and their amplitudes for the applied pulse and cavity bandwidths. These are presented as yellow vertical bars in Fig.~\ref{fig:echo}(b). 

The decay of the mean envelope of the oscillations can be fitted to a bi-exponential function [orange line in fig.~\ref{fig:echo}(a)], with a short decay time of \mbox{T$_2^{\text{(short)}}=4.50\pm0.03$~$\mu$s}, and a long decay time of \mbox{T$_2^{\text{(long)}}=109\pm2$~$\mu$s}.   As all the ESEEM modulation frequencies are equal to or higher than 366 kHz (the fundamental nuclear Zeeman frequency), the initial decay at a rate of $(2\pi\cdot4.5\ \mu s)^{-1}\approx35$~kHz cannot come from ESEEM.  
Thus, to explain the shape of the decay curve and its temperature dependence (see below), we consider three main dephasing mechanisms~\cite{Dikarov2016flipflop}: direct flip-flop of neighboring spins within the sub-ensemble probed by the cavity; instantaneous diffusion dephasing due to the flipping of neighboring spins by the $\pi$-pulse; and stochastic energy shifts (`spectral diffusion') of the probed spins due to flip-flops of the entire ensemble~\cite{Rancic2022spectralDiffusion,Alexander2022REspinDynamics}. We neglect dephasing of the electronic spins due to nuclear spins ($^{25}$Mg and $^{61}$Ni), as the product of their density and magnetic moments is much lower than that of the Ni$^{2+}$ electron spin ensemble. We do include the dephasing of the $^{25}$Mg nuclear spins themselves, as will be elaborated on below. \color{black}

At low temperatures, the first two electron spin dephasing processes usually dominate, as they involve resonant dipole-dipole interaction between close-by spins. However, in an inhomogeneous ensemble, some of the probed spins will have fewer probed-spin neighbors than others, leading to a distribution of decay times and to a bi-exponential decay curve~\cite{Shankar2010biexponent}. In general, in the sub-ensemble of probed spins for which the immediate environment is of low density, the spectral diffusion dephasing would have a more significant contribution to the total dephasing rate. However, at low temperatures the first two processes only weakly depend on temperature~\cite{Dikarov2016flipflop}, while the spectral diffusion dephasing, which depends on the number of spin pairs that can perform flip-flop, vanishes at low temperatures~\cite{Rancic2022spectralDiffusion}. Thus, at very low temperatures, the short decay time is caused by the sub-ensemble of probed spins with strong instantaneous diffusion and direct flip-flop, and the long decay time originates from the sub-ensemble in which these interactions are weak. One can model the low-temperature spin-echo trace by multiplying the calculated ESEEM trace by the fitted bi-exponential decay. As the nuclear spins also dephase (due mostly to static inhomegeneities~\cite{Mims1965ESEEM}), one has to introduce a decaying envelope term also to the oscillation visibility (see Appendix~\ref{app:ESEEM}). The result of this model is presented in Fig.~\ref{fig:echo}(a) by the solid black line. The extracted $^{25}$Mg nuclear spin inhomogeneous dephaseing time is \mbox{T$_2^{*\text{(nuc)}}=52\pm$2~$\mu$s}.\color{black}

\begin{figure}[tb]
\centering
\includegraphics[width=1\columnwidth]{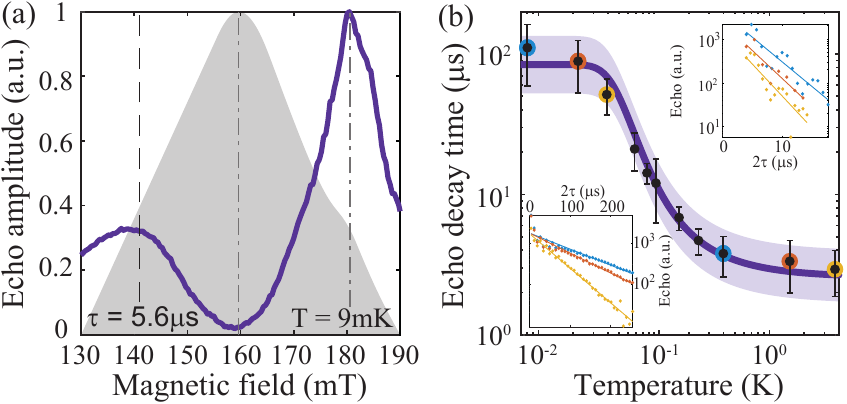}
\caption{\label{fig:decays} Dependence of spin coherence on density and temperature. (a) Spin echo amplitude (normalized to its maximum value) at $\tau=5.6$~$\mu$s and a temperature of 9~mK versus the applied magnetic field. The shaded gray area is the integrated ESR signal, proportional to the defect density probed at each magnetic-field value. The dash-dotted gray lines mark the peaks of the Ni$^{2+}$ and Cr$^{3+}$ distributions. The black dashed line marks the field at which all time-dependent echo measurements were performed. (b) Time dependence of the echo signal vs. temperature. The dots are the echo-envelope long decay times, extracted by exponential fits to the data. The error bars are the 68\% confidence level of the fits. The solid line is a two-parameter fit of the measured times to a model including the temperature-dependent, spectral-diffusion dephasing rate, and an additional, temperature-independent rate accounting for the direct flip-flop and the instantaneous diffusion processes (see Appendix~\ref{app:SpectDiff}). The shaded purple area marks the 68\% confidence interval of this fit. The bottom-left (top-right) inset presents the measured echo signals versus the time delay (dots), for the three lowest (highest) temperatures, together with single-exponential functions fitted to the long-time range (solid lines). The colors correspond to the temperature, as marked in the main plot by the solid circles.}
\end{figure}

The low-temperature short-time coherence strongly depends on the average density of the probed sub-ensemble ~\cite{Alexander2022REspinDynamics}. This can be observed by scanning the field across the inhomogeneous broadening of the spin ensemble and probing the spin-echo amplitude for a fixed, short time difference (here 5.6~$\mu$s), as shown in Fig.~\ref{fig:decays}(a). It is clearly seen that the amplitude drops near the center of the distribution, where the density of probed spins is the highest, and thus most of the probed population would experience strong direct dephasing. The strong coherence peak at 180~mT is due to Cr$^{3+}$ ions, the density of which is much lower than that of the Ni$^{2+}$ ions. 

As the temperature increases, the spectral diffusion rate increases, first affecting only the lower density sub-ensemble, until at a certain temperature it will dominate even over the direct processes in the denser sub-ensemble, at which point the coherence decay will become mono-exponential. 
Figure ~\ref{fig:decays}(b) presents the extracted long decay times versus temperature. The line is a two-parameter fit to a model including both a temperature-independent, relatively weak component, due to instantaneous diffusion and direct flip-flop in the low-density sub-ensemble, and the temperature-dependent, spectral diffusion dephasing rate~\cite{Rancic2022spectralDiffusion,Alexander2022REspinDynamics}, the latter adapted to a spin-1 bath (see Appendix~\ref{app:SpectDiff}).

The spin decoherence rate saturates around liquid-helium temperature, yielding a coherence time of about 3~$\mu$s. At these temperatures, the coherence decay is indeed mono-exponential [see top right inset to Fig.~\ref{fig:decays}(b)]. As the rates of all the three dephasing processes we consider depend at-least linearly on the ensemble density~\cite{Rancic2022spectralDiffusion}, there is a prospect of considerably prolonging the coherence time by using less dense ensembles. This, in combination with thermal annealing, could also considerably decrease the inhomogeneous broadening, bringing that of the $\delta m_s=2$ transition to the level of a few MHz, which would enable dynamical decoupling of the entire ensemble using nanosecond microwave or optical pulses. Note that this requirement could be considerably alleviated if picosecond or even femtosecond optical pulses could be used (see Sect.~\ref{sect:proposal} below). The ultimate limit is the spin lifetime, T$_1$, measured to be 1~ms at 3.5~K for a spin splitting of 9.2~GHz~\cite{Lewis1967T1,Lewis1967phononESR}. As, up to about 20~K, T$_1$ increases quadratically when decreasing the spin splitting~\cite{Lewis1967T1,Lewis1967phononESR}, tens to hundreds of milliseconds may be within reach even for these rather high temperatures. \color{black}

\section{Optical spectroscopy}\label{sect:opt_spect}
\subsection{Emission spectrum}
\begin{figure}[tb]
\centering
\includegraphics[width=1\columnwidth]{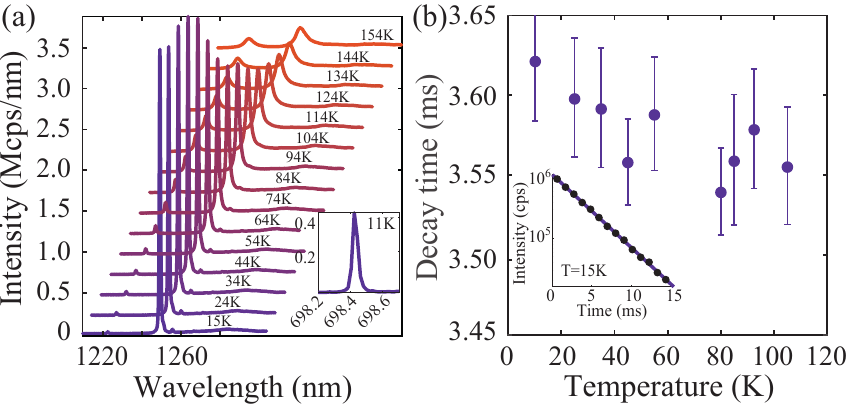}
\caption{\label{fig:opt_spectra} Optical spectroscopy of Ni$^{2+}$:MgO. (a) Temperature dependence of the emission spectrum. The inset (using the same units as in the main figure) shows the Cr$^{3+}$ line around 698~nm. (b) Fluorescence decay-time versus temperature. The inset presents a characteristic fluorescence decay curve (semi-logarithmic scale). }
\end{figure}

For optical fluorescence spectroscopy, we placed the sample in a closed cycle, low-vibration helium flow cryostat (ARS CS204-DMX-20-OM). A diode laser at 660~nm (Thorlabs LP660-SF50) was used for excitation (into a vibrational side-band of the $^3$T$_{1g}$ multiplet). The fluorescence was collected using an infrared-optimized, NA=0.8 objective lens (Shibuya M ePLAN NIR 100A) and analyzed by a 0.75~m spectrometer (Teledyne-Princeton Instruments SpectraPro HRS-750), equipped with a 300 g/mm grating (resolution limit of 30~GHz around 1250~nm), and a liquid-nitrogen-cooled InGaAs CCD array detector (Teledyne-Princeton-Instruments PyLoN IR). Figure~\ref{fig:opt_spectra}(a) presents the measured emission spectra at different temperatures. Two distinct lines, at 1220~nm and 1250~nm dominate the spectrum,  corresponding to the optical transitions from the first two excited states to the ground state. The (inhomogeneous) width of the lines up to temperatures of about 60~K is $\sim$100~GHz, much narrower than the splitting between them, 5.28~THz. While the lines further broaden at higher temperatures (mostly homogeneously), it is clearly seen that they remain well-separated up to temperatures as high as 150~K. The inset presents the emission spectrum around 698~nm, detected by the same spectrometer using a 1200 g/mm grating and a silicon CCD array detector (Teledyne-Princeton-Instruments Blaze HR). The narrow emission line (18~GHz, close to the resolution limit of 15~GHz) of Cr$^{3+}$~\cite{Imbusch1964TemperatureMgO} is clearly seen, supporting the identification of the ESR line at 180~mT (Fig.~\ref{fig:ESR}). By modulating the laser current and gating the CCD camera accordingly, we measured the time dependence of the fluorescence following the laser pulse, and confirmed that the 1250~nm fluorescence decay time in our sample is indeed $\sim$3.6~ms, as previously reported for Ni$^{2+}$ in MgO~\cite{Iverson1980NiMgOLaserParameters}. Figure~\ref{fig:opt_spectra}(b) presents the measured lifetime of the excited level versus the temperature. The inset presents an exemplary measurement. Very little change ($\sim$2\%) of the decay time is observed even up to 100~K~\cite{Iverson1980NiMgOLaserParameters}. 

These features suggest that if one of the two excited electronic levels leading to the observed emission contains a SO-coupled state, the optical coherence time would not pose a limitation on the fidelity of optical spin manipulation performed using pulses of suitable duration (shorter than the optical coherence time, longer than the SO coupling time) even at high temperatures (as long as the spectral width of the lines is smaller than the separation between them). We use magneto-optical spectroscopy measurements to verify that such a state indeed exists.

\subsection{Magneto-optical spectroscopy}
\begin{figure*}[tbh]
\centering
\includegraphics[width=0.99\textwidth]{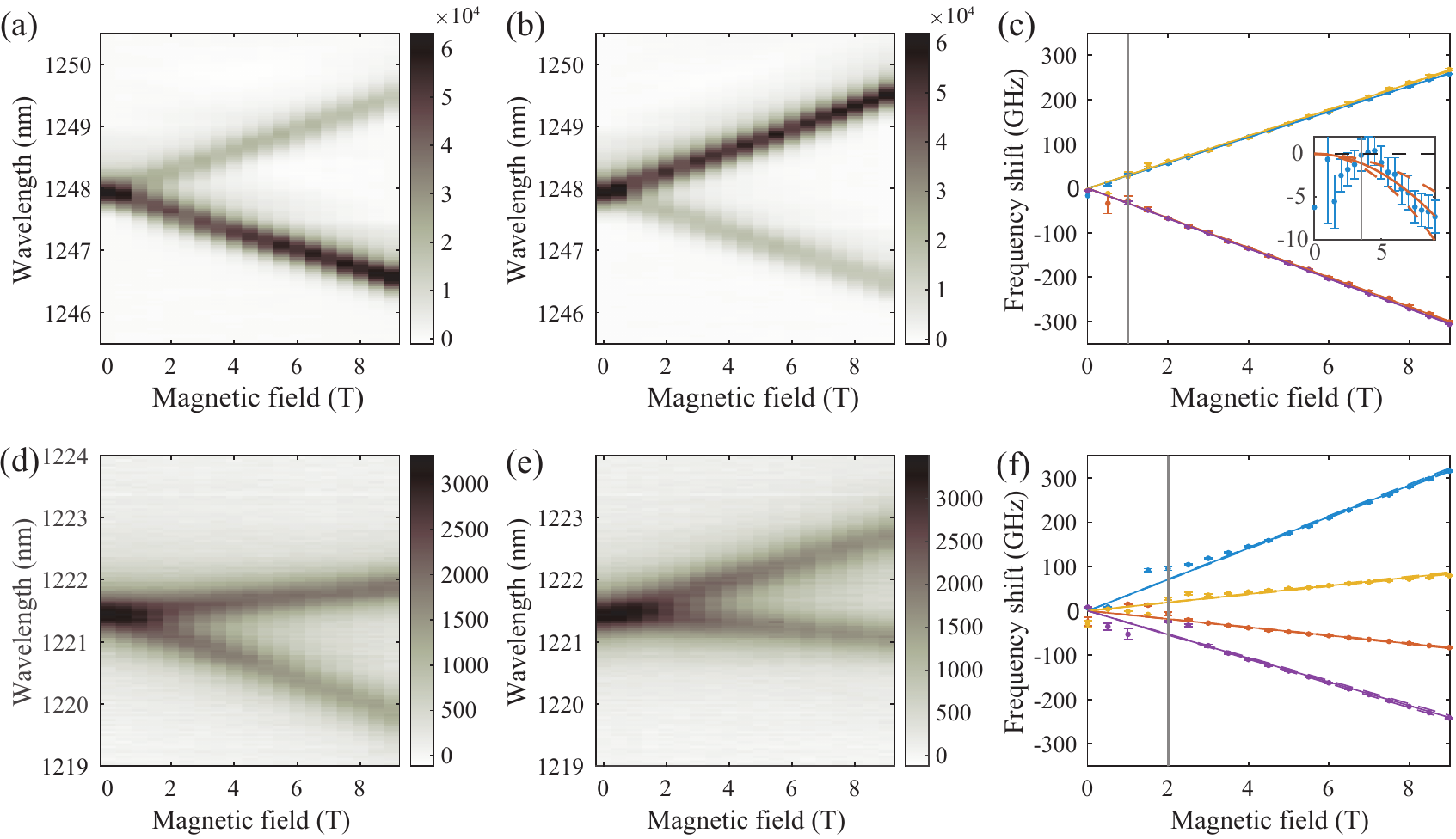}
\caption{\label{fig:magnetospect} Polarized magneto-fluorescence spectroscopy for determining spin-orbit coupling in the excited levels of Ni$^{2+}$:MgO. (a) [(b)] Measured spectra around 1250~nm for $\sigma^+$ [$\sigma^-$] circular polarization versus the applied magnetic field. (c) The line centers (dots) extracted from (a) and (b), together with linear regression fits (lines). The error bars present the errors in the line centers. For cases where only one spectral peak could be identified, only one point per data set is presented. The data points to the left of the vertical gray line were excluded from the linear regression. (d)-(f) The same as (a)-(c), for the 1220~nm emission. The inset in (c) presents the mean difference between the frequencies of oppositely-polarized transitions of the same sign of the Zeeman shift versus the magnetic field $B$. At high fields, a significant difference from zero is seen. The solid line is a fit of $\beta_\mathrm{meas} B^2$ excluding the points to the left of the gray vertical line.}
\end{figure*}

For performing polarized magneto-fluorescence spectroscopy, we placed the sample in a closed-cycle helium cryostat (attocube attoDRY 2100), equipped with a 9-T superconducting magnet. The sample temperature could be varied from 1.7~K up to room temperature, independently of the magnet temperature, which was kept low and constant. A Ti:Sapphire laser (Sirah Matisse CS) was used for optical excitation at 690~nm. Two sets of measurements were performed, one at 1.7~K base temperature, and the other one at 60~K. In each set, the magnetic field was varied from 0 to 9~T and the emission spectrum was measured in two orthogonal circular polarizations. The emission was dispersed using the HRS-750 spectrometer, equipped with a 600 g/mm grating (resolution limit of 15~GHz around 1250~nm), and recorded using an electrically-cooled CCD array camera (Teledyne-Princeton Instruments NIRvana HS). The first set focuses on the 1250~nm line (emission from the $E_u$ level) and the second on the 1220~nm line (emission from the $T_{1u}$ level). The 1220~nm emission was measured at an elevated temperature as it is extremely weak at lower temperatures, due to thermalization to the lowest excited state [see also Fig.~\ref{fig:opt_spectra}(a)]. Figure~\ref{fig:magnetospect} presents the polarized spectra for the two transitions.

At a high enough magnetic field, the 1220~nm emission splits into four, fully polarized lines, and the 1250~nm emission splits into two, partially polarized lines, with a polarization degree (defined as the ratio of the difference between the intensities in the two polarizations to their sum) of about 50\%. These patterns exactly match the prediction for the weak DJT case. In that case, the theory also predicts the Zeeman shifts of all emission lines (Appendix~\ref{app:pol_sel_rules}). 

In order to compare our measurements to the predicted Zeeman shifts, we fitted each of the measured fluorescence spectra to a double hyperbolic-secant function and extracted the energies of the two peaks for every value of the applied magnetic field. These energies are presented in Figs.~\ref{fig:magnetospect}(c) and \ref{fig:magnetospect}(f). We then fitted the magnetic field dependence of the peak energies to straight lines with a common origin. The best-fitted lines are also presented in the figures.

For the 1250~nm emission, the difference between $\sigma^+$ and $\sigma^-$ polarized lines of opposite-sign slopes is predicted to be equal to the splitting of the ground state. That is, by dividing the slope difference by $2\mu_\mathrm{B}$, one should obtain $g_s^\mathrm{g}$. In this way, we obtain a value of $g_s^\mathrm{g}=2.242\pm0.003$. This value indeed closely matches the value directly measured using ESR. The theory predicts that the two $E_u$ states should not split in a magnetic field, to first order. However,  due to the (very small) magnetic-field-induced mixing of the $E_{u,\epsilon}$ state with the $T_{1u,0}$ state, there should be a negative quadratic shift of its energy with the magnetic field~\cite{Armelles1984MagnetoSpectV3GaAs}. Using second-order perturbation theory, the coefficient of this quadratic shift is given by $\beta=-(g_s^\mathrm{e}+\tilde{g}_L)^2\mu_\mathrm{B}^2/\Delta_{21}$, where $\tilde{g}_L=g_Le^{-\kappa/2}$, and $\Delta_{21}$ is the zero-field splitting between the $E_u$ and $T_{1u}$ states. Using the measured values of $\Delta_{21}=5.28\pm0.02$~THz and $g_s^\mathrm{e}+\tilde{g}_L=1.33\pm0.02$ (obtained from the Zeeman splitting of the $T_{1u}$ states, as shown below), we obtain $\beta_\mathrm{calc}=-66\pm2$~MHz/T$^2$.
While this effect is very small, we could directly measure it by looking at the difference between the energies of $\sigma^+$ and $\sigma^-$ polarized lines having same-sign Zeeman shifts. At high enough magnetic fields we could measure a significant ($>3\sigma$) deviation from zero, as presented in the inset of Fig~\ref{fig:magnetospect}(c). It is also clear that the dependence of this deviation on the magnetic field is non-linear. The solid line is a quadratic fit to the points for which $B\geq3.5$~T. The dashed lines present the 68\% confidence level of this fit (mostly due to the uncertainty in the measured energy differences). From the fit, we obtain a value of $\beta_\mathrm{meas}=-90\pm35$~MHz/T$^2$, which agrees with the calculated value to within the measurement error. A similar effect, though much larger, was previously observed for V$^{3+}$ ions in GaAs~\cite{Armelles1984MagnetoSpectV3GaAs}. 

For the 1220~nm emission, the slope difference between the two outer lines is again only due to the ground state and is again predicted to be $2\mu_\mathrm{B}g_s^\mathrm{g}$. We indeed extract a value of $g_s^\mathrm{g}=2.22\pm0.025$, which is consistent with the value extracted from the $E_u$ lines and with the ESR value. In addition to the linear slope, there should also be a small quadratic shift, common to both lines, due to the small magnetic coupling between the $T_{1u,0}$ and the $E_{u,\epsilon}$ states. The magnitude of this effect should be the same as for the $E_{u,\epsilon}$ state, but its sign should be opposite, that is, we expect a positive quadratic shift. However, as the signal here was much weaker than in the 1250~nm line, the accuracy of the assigned energies was lower, and we could not resolve this effect. 

The slope difference between the two inner lines is related only to the splitting of the excited $T_{1u}$ states and is predicted to be $\alpha=\mu_\mathrm{B}(\tilde{g}_L+g_s^\mathrm{e})$. Assuming that the deviation of the electron g-factor from the vacuum electron g-factor, $g_0=2.0023$, is due to SO mixing alone (that is, neglecting the crystal field contribution),
and taking into account only the closest $T_{1u}$ state (which is that arising from the $^3$T$_1$ manifold), one obtains $g_{s,\mathrm{th}}^\mathrm{e}=1.84$ (Appendix~\ref{app:T1_gfactor}).
Together with the theoretical value of $\tilde{g}_{L,\mathrm{th}}=-0.47$ (see Appendix~\ref{app:DJTmodel} and Appendix~\ref{app:gL}), we obtain $\alpha_{\mathrm{th}}=1.37\mu_\mathrm{B}$. This predicted value is in a good agreement with the measured value of $\alpha_{\mathrm{meas}}=(1.33\pm0.02)\mu_\mathrm{B}$. An even better agreement may be obtained if the effects of the crystal field on the g-factor would be taken into account~\cite{Misetich1964gfactor}. 

\section{Proposed protocols}\label{sect:proposal}
Having established that the $^3$T$_2$ excited states of Ni$^{2+}$ in MgO involve unquenched, THz level SO coupling, and can thus mediate fast spin-photon coupling, in this section we present several  protocols for optical control of the ground-state electron spin.  
\subsection{Spin state preparation and measurement}
\begin{figure*}[tbh]
\centering
\includegraphics[width=0.99\textwidth]{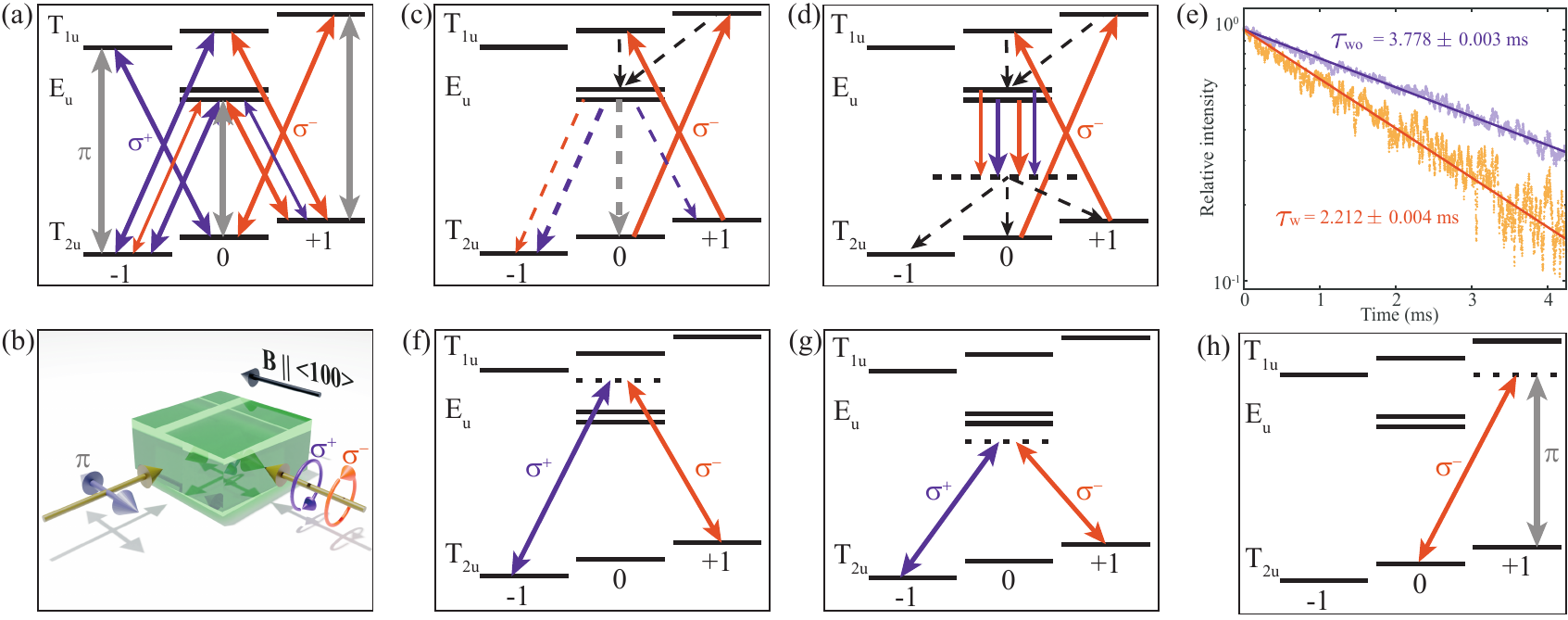}
\caption{\label{fig:basic_schemes} Optical spin preparation, measurement, and control schemes. (a) The polarization selection rules for optical transitions between the ground state manifold, $T_{2u}$, and the first two excited state manifolds, $E_u$ and $T_{1u}$. Purple, orange, and grey arrows represent transitions in $\sigma^+$, $\sigma^-$, and $\pi$ polarizations, respectively. The widths of the arrows represent the strength of the transitions (see also Table~\ref{tab:weak_sr} in Appendix~\ref{app:pol_sel_rules}). (b) The geometry of exciting optical beams with respect to the crystal axes and the applied magnetic field. (c) Polarized optical pumping to the $T_{2u,-1}$ state. Dashed purple, orange, and grey (black) lines indicate spontaneous (non)radiative decay. Similar schemes with different pump polarizations allow pumping to the other two ground states. (d) Pulsed state preparation. Here, the pumping is done using $\pi$-pulses (of $\sigma^-$-polarized light, for example) that transfer all undesired state population to the upper excited state. From there, all this population quickly decays to the lower excited state. It can either be shelved there for the lifetime of the lower excited state or could be forced down to a phonon side band, which then quickly decays to the ground state. In the latter case, to achieve full spin-state polarization, the process has to be repeated a few times. (e) Fluorescence decay curves without (purple) and with (orange) the introduction of laser light at the phonon side-band. A clear increase in the decay rate is observed. (f) [(g)] Coherent optical arbitrary spin rotation on the $\{T_{2u,1},T_{2u,-1}\}$ qubit space using a polarized pulse around the $T_{1u}$ [$E_u$] transition. (h) Coherent optical arbitrary spin rotation on the $\{T_{2u,1},T_{2u,0}\}$ qubit space.}
\end{figure*}

Figure \ref{fig:basic_schemes}(a) presents the polarization selection rules between the ground-state manifold and the two lowest excited-state manifolds, where the definitions of the polarizations with respect to the magnetic field and crystal axes are presented in Fig.~\ref{fig:basic_schemes}(b) (see also Table~\ref{tab:weak_sr} in Appendix~\ref{app:pol_sel_rules}). These selection rules allow for polarization-based spin state preparation. Figure~\ref{fig:basic_schemes}(c) presents the basic principle: excitation with a defined polarization leaves one of the three ground state spin sub-levels uncoupled to the excitation field. If the lifetime of this state is longer than the decay time from the excited state, most of the population will eventually accumulate in this state. 

While the nonradiative decay from the $T_{1u}$ to the $E_u$ excited states is fast even at low temperature [as evident from the thermalization of the excited state population, Fig.~\ref{fig:opt_spectra}(a)], the decay from the $E_u$ excited state to the $T_{2u}$ ground states takes a few milliseconds, even at high temperatures [Fig.~\ref{fig:opt_spectra}(b)]. As the ground-state spin lifetime has been measured to be on the orders of milliseconds only at temperatures of a few Kelvin~\cite{Lewis1967T1}, the standard optical pumping, Fig.~\ref{fig:basic_schemes}(c), may not work at higher temperatures.

There are at least two solutions to this problem. One is to use the fact that the $E_u$ state is long-lived and use it as a shelving state that stores the unwanted spin population while coherent operations are performed on the population that remains in the ground state, which is only of the desired spin state. For this, the polarized optical field should transfer all the unwanted population to the excited state before it decays. This can be achieved using an ultra-fast, optical $\pi$-pulse, resonant with the 1220~nm transition ($T_{2u}\rightarrow T_{1u}$). In order to transfer the entire population, the excitation has to be coherent. This means that the pulse bandwidth has to be much larger than the optical line width ($\sim100$~GHz). However, in order to not involve the $E_u$ levels, which will spoil the polarization selection rules, the pulse spectrum has to be narrower than the energy difference between the $E_u$ and $T_{1u}$ spectral lines (5.28~THz). A $\sim1$~THz wide pulse, that is, of a few hundred fs duration, would fit this range. One downside of this solution is that one decreases the optical density (the effective defect density) by a factor of three (as only a third of the defects are left in the ground state). Further, the shelved population is in a random spin state, making it a source of spectral diffusion noise. \color{black}

The second solution may overcome these two issues. As shown in Fig.~\ref{fig:basic_schemes}(d), a second pulse, introduced after all the excited population has decayed into the $E_u$ state, at a frequency matching the transition from the shelving $E_u$ state to a phonon side-band of the ground state, would stimulate the transition of the shelved population to the phonon side-band. From there, the population would quickly decay back to the ground state. Repeating the shelving and stimulating pulses a few times would result in most of the population being pumped into the ground state decoupled from the shelving pulse, in a similar manner to standard optical pumping. 

To test the feasibility of this `stimulated optical pumping' concept, we introduced about 0.5~W of laser light at 1319~nm (Sanctity Laser SSL-1319-1500-10TM-D-LED) during the decay of the population after its excitation by the 660-nm pulse. Figure~\ref{fig:basic_schemes}(e) presents the fluorescence decay with and without the addition of the 1319~nm laser. A clear decrease in the fluorescence decay time is observed, indicating the stimulation of population decay from the shelving state. Using a CW laser, however, is inefficient, as only a small part of the phonon side-band is used, and the added power heats up the sample. Using an ultra-short pulse for the stimulated de-excitation should solve these issues. 

For optically measuring the spin state, one can turn on again the polarized pumping light at 1220~nm and monitor the resulting fluorescence at 1250~nm. As for each of the three polarizations ($\sigma^+$, $\sigma^-$, and $\pi$), one of the three spin states is uncoupled from the polarized pumping field, the fluorescence intensity will be inversely proportional to the population of that state. The combined information from all three measurements would thus yield the populations of all three states. 

\subsection{Coherent spin control}
The polarized $\Lambda$-systems present in the level structure allow for polarization-based, full coherent control of the spin states using short optical pulses. Figures \ref{fig:basic_schemes}(f) and \ref{fig:basic_schemes}(g) present the relevant transitions for coherent control of the $\{T_{2u,1},T_{2u,-1}\}$ qubit manifold, through the $T_{1u,0}$ state or the $E_u$ states, respectively. In the latter case, the coupling paths through the two $E_u$ states destructively interfere only partially, still enabling control. In both cases, the control pulse can be off-resonance or near-resonance, whereas in the latter case the pulse spectrum should be much wider than the line-width of the relevant transition. The axis of rotation in the Bloch sphere is determined by the pulse polarization. The angle of rotation about this axis is determined either by the intensity of the pulse (off-resonant pulse)~\cite{Press2088OptContQDRaman} or by the detuning of the pulse (near-resonant pulse)~\cite{Poem2011OpticallyDot}; in the 
latter case, the intensity is set to create a full $2\pi$ rotation, ending back at the ground state~\cite{Kodriano2012CompletePulse}. Figure~\ref{fig:basic_schemes}(h) shows the transitions employed to control the $\{T_{2u,1},T_{2u,0}\}$ qubit manifold. As they involve both $\sigma$ and $\pi$ polarizations, they have to be applied from orthogonal directions [Fig.~\ref{fig:basic_schemes}(b)]. A similar arrangement with the opposite $\sigma$ polarization would drive the $\{T_{2u,0},T_{2u,-1}\}$ qubit manifold. Here too, both near- and off-resonance control pulses may be applied.

\subsection{Noise-free quantum memory}
The ability to optically prepare and coherently control their spin state, combined with the near-telecom optical transitions and the possibility of creating high optical-density ensembles, naturally suggests the application of divalent nickel ion ensembles in MgO as quantum-optical memories~\cite{Simon2010QuantumMemoryReview,Heshami2016QuantumMemoryReview}.  The quantum memory scheme most suitable to a medium with a large broadening of the excited state is the far-detuned Raman scheme~\cite{Nunn2007RamanMemTheory}. The optical cooperativity $\mathcal{C}$ of the system, which governs the memory efficiency~\cite{Poem2015BroadbandDiamond}, can be estimated using the transition dipole moment of the $T_{2u}\leftrightarrow T_{1u}$ transition, $3.4\times10^{-32}$~Cm, as derived from the oscillator strength of $4\times10^{-7}$~\cite{Pappalardo1961OpticalI}. We assume a density of $2.4\times10^{18}$~cm$^{-3}$ (10 times lower than the density of the current sample), a detuning of 200~GHz, and a wave-guide of 5~mm in length and a cross-section of 5$\times5$~$\mu$m$^2$. For control pulses of 1~$\mu$J, which are readily produced by standard optical parametric amplifiers, we obtain $\mathcal{C}\approx 2$, indicating a total storage and retrieval efficiency of $\mathcal{C}^2/(1+\mathcal{C})^2\approx 45\%$~\cite{Nunn2007RamanMemTheory}. Thus, it seems that efficient Raman storage should be possible using realistic parameters. Furthermore, much higher cooperativity, $\mathcal{F}\times\mathcal{C}$, and hence higher efficiency, can be obtained by adding an optical cavity with moderate finesse $\mathcal{F}$ \cite{Lahad2017InducedCavities,Nunn2017TheoryCavity,Saunders2016Cavity-EnhancedMemory}, \textit{e.g.}, by using a ring-resonator structure or imprinting Bragg mirrors onto the wave-guide. 

One prevalent source of noise in a Raman memory scheme is due to four-wave mixing~\cite{Michelberger2015SinglePhotonRamanMemory}. This is the case when the control field couples to the full ground-level manifold, and not only to the empty ground state, a situation aggravated at detunings larger than the ground-state splitting. However, in a Raman memory based on a polarized $\Lambda$ system, a polarized control field couples only to a single ground state, and thus four-wave-mixing noise is suppressed~\cite{Poem2015BroadbandDiamond,Treutlein2022PolMem}. 

\begin{figure*}[tb]
\centering
\includegraphics[width=0.99\textwidth]{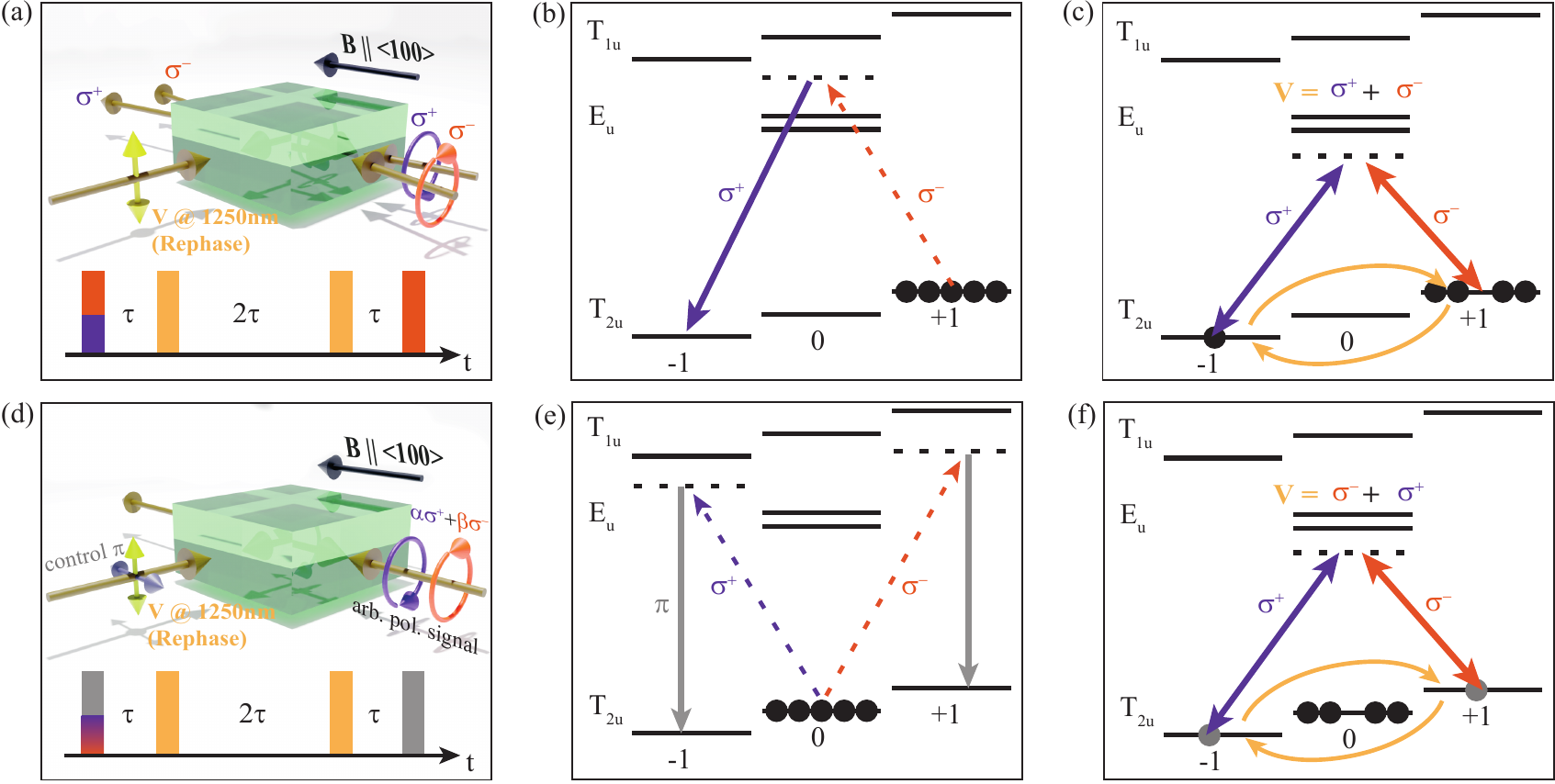}
\caption{\label{fig:memory_schemes} Noise-free memory schemes. (a)-(c) A single-mode, optionally re-phased memory scheme. (a) Spatial configuration (top) and timing (bottom) of the optical pulses. (b) Storage of a $\sigma^+$ signal by creating a $T_{2u,1}-T_{2u,-1}$ coherence in a medium initially prepared in the $T_{2u,1}$ ground state, by a $\sigma^-$ control field. A second application of the control before the inhomogeneous dephasing time of the $T_{2u,1}-T_{2u,-1}$ coherence would read the signal out. (c) A pair of re-phasing pulses of vertical polarization ($V$) can be applied between storage and retrieval to cancel out the inhomogeneous dephasing, thus prolonging the memory time. (d)-(f) A polarization-conserving, optionally re-phased memory scheme. (d) Spatial configuration (top) and timing (bottom) of the optical pulses. (b) Storage of an arbitrarily-polarized signal on the pair of coherences, $T_{2u,0}-T_{2u,1}$ and $T_{2u,0}-T_{2u,-1}$, using a $\pi$-polarized pulse on a medium initially prepared in the $T_{2u,0}$ ground state. A second application of the control before these coherences decay would read the signal out. (f) Here too, a pair of re-phasing pulses of $V$ polarization can be applied.  }
\end{figure*}

A second source of noise is the leakage of control light into the signal channel. When the signal and control are oppositely polarized, they can be separated by their polarizations. Usually, however, this is not enough, and a second separation mechanism, such as spectral filtering is invoked~\cite{Reim2011RamanMemory,Michelberger2015SinglePhotonRamanMemory,Saunders2016Cavity-EnhancedMemory,Thomas2019NoiseFreeRamanAbsorption,Davidson2022ArxivMemory}. This would be possible here only if the spectrum of the control pulse is narrower than the ground-state splitting, but that would limit the bandwidth of the memory. Furthermore, tight spectral filtering usually lowers the efficiency of the memory. Here we propose to replace spectral filtering with spatial filtering, by introducing an angle between the signal and control modes. This is possible as, in contrast to warm atomic vapors, here the emitters do not move during storage and cannot create any motional dephasing due to signal and control wave-vector mismatch~\cite{Finkelstein2019Narrowing,finkelsteinPRX2020}. Figure.~\ref{fig:memory_schemes} presents two polarized Raman-memory schemes with an additional spatial-mode mismatch, applicable in the Ni$^{2+}$:MgO system.

The first scheme, described in the top three panels of Fig.~\ref{fig:memory_schemes}, uses the $\sigma^+$-polarized transition $T_{2u,1}\leftrightarrow T_{1u,0}$ for the signal, and the $\sigma^-$-polarized transition $T_{2u,-1}\leftrightarrow T_{1u,0}$ for the control, where the ensemble is first prepared in the $T_{2u,1}$ state. As shown in Fig.~\ref{fig:memory_schemes}(a) (top), while the control field propagates along the magnetic field (setting the quantization axis), the signal field is at a small angle with respect to it. As shown in Fig.~\ref{fig:memory_schemes}(b), this scheme is based on the $\{T_{2u,1},T_{2u,-1}\}$ qubit system. As shown in Fig.~\ref{fig:ESR}, while the relevant spin transition is narrower than that of the $\{T_{2u,1},T_{2u,0}\}$ qubit system, it is still considerably broadened. However, most of this broadening is due to inhomogeneous strain distribution, which could be mitigated by using a spin-echo sequence (and, generally, would be narrower in annealed samples~\cite{Smith1969EPRMgO}). Figure~\ref{fig:memory_schemes}(c) presents a possible way to introduce the echo pulses optically. This could be done using a vertically-polarized pulse near the $E_u$ resonance, flipping between the $T_{2u,1}$ and $T_{2u,-1}$ states. As shown in Fig.~\ref{fig:memory_schemes}(a) (top), this pulse could potentially be applied from the side of the sample, perpendicular to the control and the signal, to minimize scattering into the signal mode, but it could also be applied along their direction, as it could rather easily be spectrally filtered-out. 

Figure~\ref{fig:memory_schemes}(a) (bottom) presents a possible pulse scheme including two re-phasing pulses between storage and retrieval, where the time between the memory control pulses and the re-phasing pulses is $\tau$, and the time between the two re-phasing pulses is $2\tau$. This scheme would return the qubit state to its initial state in time for retrieval. This is similar to the revival of silenced echo (ROSE) quantum memory scheme~\cite{Damon2011ROSE}, as in the absence of a microwave cavity, the spin echo between the two re-phasing pulses would be very weak. \color{black}

Figure~\ref{fig:memory_schemes}(d) presents a second noise-free memory scheme. This scheme, in contrast to the first one, stores both polarization modes of the signal. As shown in Fig.~\ref{fig:memory_schemes}(e), Starting from the system initialized in the $T_{2u,0}$ state, a $\pi$-polarized control pulse stores an arbitrarily $\sigma^\pm$-polarized signal in the $T_{2u,0}-T_{2u,1}$ and the $T_{2u,0}-T_{2u,-1}$ coherences. The $\pi$-polarized control pulse is introduced from the side, so, despite being in the same frequency as the signal, can be separated by both polarization and spatial filtering. As the storage coherences include the broad $T_{2u,0}$ state, here re-phasing is critical. Fortunately, the same two-pulse re-phasing scheme described above can be used here as well. As the $T_{2u,1}$ and $T_{2u,-1}$ states are switched, the phase dispersion of both $T_{2u,0}-T_{2u,1}$ and $T_{2u,0}-T_{2u,-1}$ coherences is reversed [Fig.~\ref{fig:memory_schemes}(f)]. One downside of this scheme is that, as the control comes from the side, the area it has to illuminate is much larger, necessitating a much larger pulse energy. This may be mitigated by embedding the waveguide into a planar micro-cavity, resonant with the control frequency, effectively enhancing the control power acting on the storage medium. \color{black}

For the two memory schemes proposed above, using two re-phasing pulses limits the applicability of such memories to cases where the required storage time is known in advance. This is the case for the synchronization of random events to fixed, pre-determined time bins. 
Nevertheless, the memory schemes could be adapted to cases where the release time-bin is not pre-determined. This can be done by setting the total re-phasing time, $4\tau$, to the minimum cycle time of the synchronized system, repeating the two-pulse re-phasing sequence until retrieval is required, and then introducing the retrieving control at the end of the last re-phasing cycle. This would amount to applying a Carr-Purcell-Meiboom-Gill (CPMG) sequence of variable length between storage and retrieval, and would thus have the added value of protecting the stored coherence from dynamical external noise (up to a frequency of $1/4\tau$). 
Moreover, the applied pulse sequence does not have to be limited to CPMG. It could be any other periodic pulse sequence that can be terminated after an arbitrary number of periods. For example, one could apply a combination of the CPMG sequence and the Waugh-Huber-Haeberlen (WAHUHA) sequence~\cite{Farfurnik2018WAHUHA}, which protects both against external dynamical noise and against noise created by dipolar interactions within the ensemble, thus prolonging the memory time beyond the limit set by dipolar dephasing. This would alleviate the limit on the density of the ensemble, potentially further increasing the memory efficiency. 

\section{Conclusion}\label{sect:conclusion}
We have introduced general guidelines for identifying electron spin defects in solid-state systems that could coherently couple to light in the presence of significant optical decoherence, may it be caused by inhomogeneous broadening in an ensemble, high temperature, or both. The two main requirements are an excited-state SO interaction faster than the optical decoherence and an orbital-singlet ground state. Following these guidelines, we propose to study transition-metal ion ensembles in various crystals, and, as an example, we present the case of Ni$^{2+}$ in MgO.

First, we perform ESR measurements of Ni$^{2+}$:MgO and extract both the inhomogeneous and homogeneous coherence times. In our sample, these are on the order of a few ns and a few $\mu$s, respectively, at liquid-helium temperature. Using temperature-dependant spin-echo measurement, we find that the main homogeneous broadening mechanism at this temperature is spectral diffusion due to dipolar interactions within the ensemble and that it is fully saturated. As the dephasing rate due to this mechanism depends on the density of the ensemble, the dephasing time could be prolonged by working with lower densities. Our particular sample is dense, 
containing $\sim 450$~ppm of Ni, and indeed lower densities would still be high enough to allow for efficient interaction with light. Alternatively, one could use designated dynamical-decoupling sequences to protect against dipolar dephasing   or to use optical pumping techniques to polarize the spin ensemble. The ultimate limit is the spin lifetime, which, by working at small spin splittings, may be made as high as hundreds of milliseconds, even at temperatures as high as 20~K, on par or even exceeding the spin liftimes of rare earth ions at these temperatures~\cite{GuptaErLifetime2023}. Consequently, for single spin centers in dilute samples, where dipolar dephasing is negligible, our results infer that long coherence times could be achieved up to temperatures of several tens of K, where low-cost, high-cooling-power Stirling coolers can be used.   \color{black}

We could also detect ESEEM, which we attributed to $^{25}$Mg nuclei in the host MgO crystal. This enabled us to extract an inhomogeneous nuclear spin coherence time of 52~$\mu$s at 9~mK. This may suggest that the homogeneous nuclear coherence time (which we have not measured in this work) may be very long, opening up the possibility of using the nuclear spin ensemble as a long-term quantum memory~\cite{London2013HHnuclearSpin,AlkaliNobleEntanglementKatz2020PRL,Bartling2022C13pair}. 

Second, we perform magneto-optical spectroscopy measurements and verify that the excited states are indeed SO-coupled states, despite a weak DJT distortion of the electronic orbitals. This, combined with the observations of spectrally-separated emission lines and an almost fixed fluorescence decay time up to higher than liquid nitrogen temperatures, suggests that this system may be used as a coherent spin-photon interface at relatively high temperatures,   limited only by the ground-state spin-coherence time. As discussed above, further work is required to measure this time versus temperature and establish the upper temperature limit. 

Third, we propose detailed protocols for optical spin-state initialization and measurement and optical coherent spin control. Specifically, even when the spin lifetime is shorter than the excited-state lifetime, spin initialization is still possible using either shelving or pulsed stimulated decay. Based on these basic protocols, we then propose two noise-free, high-bandwidth quantum memory protocols, possibly combining dynamical decoupling. While the first protocol is limited to single-mode storage, the second is for a two-mode, polarization-preserving memory. Much more work is required in order to implement these protocols and to fully explore the potential of Ni$^{2+}$ in MgO as a solid-state light-spin interface. 

Moving forward, other transition-metal dopant systems could be explored. Some interesting candidates are Co$^{2+}$~\cite{Koidl1973CoZnSweakDJT} and Ni$^{3+}$~\cite{Thurian1992ZeemanNi3ZnO} ($d^{7}$ ions), or Fe$^{6+}$~\cite{Brunold1994Fe6}, Nb$^{3+}$~\cite{Ammerlahn1997SpectroscopicGaAs} and Ta$^{3+}$~\cite{Ushakov2018IntracenterTelluride} ($d^{2}$ ions), in tetrahedral sites. All of these are known to have optical transitions between 1550 and 1700~nm, with weak or non-existing DJT quenching. In addition, many other transition-metal-ion doped crystals that match our general guidelines have not been studied at all. 

We thus believe that this work can open the way to further investigations of transition-metal ions in crystals as a new family of materials with the potential of serving as the long-sought-after high-temperature, coherent solid-state spin-photon interface, with major applications in quantum networks.

\begin{acknowledgments}
EP would like to thank H. Bernien and D. D. Awschalom for facilitating his visit to the University of Chicago and for fruitful discussions. EP was supported in part by a grant from Fran Morris Rosman and Richard Rosman. EP, LS, and OF acknowledge support from the US-Israel Binational Science Foundation (BSF) and US National Science Foundation (NSF), and the Estate of Louise Yasgour. SG and TZ acknowledge support by the National Science Foundation (NSF) Faculty Early Career Development Program (CAREER) Grant (No. 1944715), partial support by the University of Chicago Materials Research Science and Engineering Center, which is funded by the NSF under award number DMR- 2011854, and the Army Research Office grant no. W911NF2010296. IM was supported by an Alfred J. and Ruth Zeits Research Fellowship. JNB acknowledges support by the Cowen Family Endowment at MSU.   
\end{acknowledgments}

\appendix

\renewcommand{\theequation}{\Alph{section}\arabic{equation}}
\renewcommand{\thetable}{\Alph{section}\arabic{table}}

\setcounter{equation}{0}
\setcounter{table}{0}
\section{States and polarization selection rules}\label{app:pol_sel_rules}
The three ground states can be described as spin-orbit product states, between the $|$A$_{2g}\rangle$ orbital and the three spin states $|1\rangle_s$, $|0\rangle_s$, and $|-1\rangle_s$. The excited states can be described as super-positions of spin-orbit product states between the three T$_{2g}$ orbitals $|$T$_{2g,1}\rangle$, 
$|$T$_{2g,0}\rangle$, and $|$T$_{2g,-1}\rangle$, and the three spin states.

The optical transition probabilities between any excited state $|j\rangle_e$  and ground state $|i\rangle_g$ are given by
\begin{equation}
P_{ij}^k=|_e\langle j|D_M^k|i\rangle_g|^2,
\end{equation}
where $D_M^k$ is the magnetic-dipole transition matrix in polarization $k$. In the spin-orbit product basis used here, these matrices are given by
\begin{equation}
D_M^k\propto|k\rangle_l\langle \mathrm{A}_{2g}|\otimes I_s.
\end{equation}
The state $|k\rangle_l$ is the T$_{2g}$ orbital state corresponding to the polarization $k$, according to~\cite{Griffith1961}
\begin{equation}  
\{\sigma^+,\pi,\sigma^-\}\Leftrightarrow\{|\mathrm{T}_{2g,-1}\rangle,|\mathrm{T}_{2g,0}\rangle,|\mathrm{T}_{2g,1}\rangle\}.
\end{equation}
The matrix $I_s$ is a unit matrix in the spin-space, and ``$\otimes$"  is the Kronecker product. The proportionality constant is not important for the calculation of relative rates.

The ground and excited-state wave functions, calculated according to the Hamiltonian of Eq.~\ref{eq:es_SO_DJT} of the main text, the relative optical transition rates between them (in all polarizations), and the Zeeman shift coefficient for these transitions are given in Table~\ref{tab:weak_sr} and Table~\ref{tab:strong_sr} for weak and strong DJT distortion, respectively. The tables neglect second-order magnetic coupling effects.
\begin{table*}[tbh]
   \caption{Transition Zeeman shift in units of $\mu_\mathrm{B} B$ (left) and relative transition probabilities in the three polarizations, [$\sigma^+$,$\pi$,$\sigma^-$] (right), for the case of a \emph{weak} DJT distortion. The orbital $|$T$_{2g,k}\rangle$ is abbreviated by $|k\rangle_l$.}
    \label{tab:weak_sr}
\begin{ruledtabular}
    \centering
    \begin{tabular}{|c|c||c|c|c|}
        & & $|$A$_{2g}\rangle|\mathrm{-}1\rangle_s$ & $|$A$_{2g}\rangle|0\rangle_s$ & $|$A$_{2g}\rangle|1\rangle_s$ \\
       \hline
       \hline
        $E_{u,\epsilon}$ & $\tfrac{1}{\sqrt{2}}(|1\rangle_l|1\rangle_s+|\mathrm{-}1\rangle_l|\mathrm{-}1\rangle_s)$ & $g_s^\mathrm{g}$  [$\tfrac{1}{2}$ 0 0]& 0  [0 0 0]& $-g_s^\mathrm{g}$  [0 0 $\tfrac{1}{2}$]\\
        \hline
        $E_{u,\theta}$ & $\tfrac{1}{\sqrt{6}}(|1\rangle_l|\mathrm{-}1\rangle_s+|\mathrm{-}1\rangle_l|1\rangle_s)+\tfrac{\sqrt{2}}{\sqrt{3}}|0\rangle_l|0\rangle_s$ & $g_s^\mathrm{g}$  [0 0 $\tfrac{1}{6}$]& 0  [0 $\tfrac{2}{3}$ 0]& $-g_s^\mathrm{g}$  [$\tfrac{1}{6}$ 0 0]\\
        \hline
        \hline
        $T_{1u,1}$ & $-\tfrac{1}{\sqrt{2}}(|1\rangle_l|0\rangle_s+|0\rangle_l|1\rangle_s)$ & $\tfrac{1}{2}(2g_s^\mathrm{g}+g_s^\mathrm{e}+\tilde{g}_L)$  [0 0 0]& 
        $\tfrac{1}{2}(g_s^\mathrm{e}+\tilde{g}_L)$  [0 0 $\tfrac{1}{2}$]& $-\tfrac{1}{2}(2g_s^\mathrm{g}-g_s^\mathrm{e}-\tilde{g}_L$)  [0 $\tfrac{1}{2}$ 0]\\
        \hline
        $T_{1u,0}$ & $\tfrac{1}{\sqrt{2}}(|1\rangle_l|1\rangle_s-|\mathrm{-}1\rangle_l|\mathrm{-}1\rangle_s)$&  $g_s^\mathrm{g}$  [$\tfrac{1}{2}$ 0 0]& 
        0  [0  0 0]& $-g_s^\mathrm{g}$  [0 0 $\tfrac{1}{2}$]\\
        \hline
        $T_{1u,-1}$ & $\tfrac{1}{\sqrt{2}}(|\mathrm{-}1\rangle_l|0\rangle_s+|0\rangle_l|\mathrm{-}1\rangle_s)$ & $-\tfrac{1}{2}(2g_s^\mathrm{g}+g_s^\mathrm{e}+\tilde{g}_L)$  [0 $\tfrac{1}{2}$ 0]& 
        $-\tfrac{1}{2}(g_s^\mathrm{e}+\tilde{g}_L)$  [$\tfrac{1}{2}$ 0  0]& $\tfrac{1}{2}(2g_s^\mathrm{g}-g_s^\mathrm{e}-\tilde{g}_L$)  [0 0 0]\\
    \end{tabular}   
\end{ruledtabular}   
\end{table*}

\begin{table*}[tbh]
    \caption{Transition Zeeman shift in units of $\mu_\mathrm{B} B$ (right) and relative transition probabilities in the three polarizations, [$\sigma^+$,$\pi$,$\sigma^-$] (left), for the case of \emph{strong} DJT distortion. The orbital $|$T$_{2g,k}\rangle$ is abbreviated by $|k\rangle_l$.}
    \label{tab:strong_sr}
\begin{ruledtabular}
    \centering
    \begin{tabular}{|c|c||c|c|c|}
        & & $|$A$_{2g}\rangle|\mathrm{-}1\rangle_s$ & $|$A$_{2g}\rangle|0\rangle_s$ & $|$A$_{2g}\rangle|1\rangle_s$ \\
       \hline
       \hline
        $E_{u,\epsilon}$ & $\tfrac{1}{\sqrt{2}}(|1\rangle_l|1\rangle_s+|\mathrm{-}1\rangle_l|\mathrm{-}1\rangle_s)$ & $g_s^\mathrm{g}$  [$\tfrac{1}{2}$ 0 0]  & 0  [0 0 0]& $-g_s^\mathrm{g}$  [0 0 $\tfrac{1}{2}$] \\
        \hline
        $E_{u,\theta}$ & $\tfrac{1}{\sqrt{6}}(|1\rangle_l|\mathrm{-}1\rangle_s+|\mathrm{-}1\rangle_l|1\rangle_s)+\tfrac{\sqrt{2}}{\sqrt{3}}|0\rangle_l|0\rangle_s$ & $g_s^\mathrm{g}$  [0 0 $\tfrac{1}{6}$] & 0  [0 $\tfrac{2}{3}$ 0] & $-g_s^\mathrm{g}$  [$\tfrac{1}{6}$ 0 0] \\
        \hline
        $A_{2u}$ & $\tfrac{1}{\sqrt{3}}(|1\rangle_l|\mathrm{-}1\rangle_s+|\mathrm{-}1\rangle_l|1\rangle_s)-|0\rangle_l|0\rangle_s)$ & $g_s^\mathrm{g}$ 
 [0 0 $\tfrac{1}{3}$]  & 0  [0 $\tfrac{1}{3}$ 0] & $-g_s^\mathrm{g}$  [$\tfrac{1}{3}$ 0 0] \\
        \hline
        \hline
        $\tfrac{1}{\sqrt{2}}(T_{2u,1}-T_{1u,-1})$ & $|0\rangle_l|1\rangle_s$ & $g_s^\mathrm{g}-g_s^\mathrm{e}$  [0 1 0]& 
        $-g_s^\mathrm{e}$  [0 0 0]& $-g_s^\mathrm{g}-g_s^\mathrm{e}$  [0 0 0]\\
        \hline
        $T_{1u,0}$ & $\tfrac{1}{\sqrt{2}}(|1\rangle_l|1\rangle_s-|\mathrm{-}1\rangle_l|\mathrm{-}1\rangle_s)$&  $g_s^\mathrm{g}$  [$\tfrac{1}{2}$ 0 0]& 
        0  [0  0 0]& $-g_s^\mathrm{g}$  [0 0 $\tfrac{1}{2}$]\\
        \hline
        $\tfrac{1}{\sqrt{2}}(T_{1u,1}-T_{2u,-1})$ & $|0\rangle_l|\mathrm{-}1\rangle_s$ & $g_s^\mathrm{g}+g_s^\mathrm{e}$  [0 0 0]& 
        $g_s^\mathrm{e}$  [0 0 0]& $-g_s^\mathrm{g}+g_s^\mathrm{e}$  [0 1 0]\\
        \hline
        $-\tfrac{1}{\sqrt{2}}(T_{1u,-1}+T_{2u,1})$ & $|1\rangle_l|0\rangle_s$ & $g_s^\mathrm{g}$  [0 0 0]& 
        0  [1 0 0]& $-g_s^\mathrm{g}$  [0 0 0]\\
        \hline
        $T_{2u,0}$ & $\tfrac{1}{\sqrt{2}}(|1\rangle_l|\mathrm{-}1\rangle_s-|\mathrm{-}1\rangle_l|1\rangle_s)$&  $g_s^\mathrm{g}$  [$\tfrac{1}{2}$ 0 0]& 
        0  [0 0 0]& $-g_s^\mathrm{g}$  [0 0 $\tfrac{1}{2}$]\\
        \hline
        $\tfrac{1}{\sqrt{2}}(T_{1u,1}+T_{2u,-1})$ & $|\mathrm{-}1\rangle_l|0\rangle_s$ & $g_s^\mathrm{g}$  [0 0 0]& 
        0  [0 0 1]& $-g_s^\mathrm{g}$  [0 0 0]\\
    \end{tabular}
   
\end{ruledtabular}    
\end{table*}

\setcounter{equation}{0}
\setcounter{table}{0}
\section{DJT model and the absorption spectra}\label{app:DJTmodel}
The model in Eqs.~(\ref{eq:es_H})-(\ref{eq:es_H_B}) has eight independent parameters: the electron-phonon coupling energy $E_\mathrm{JT}$, the energy of the lowest phonon mode $\hbar\omega_\mathrm{ph}$, the pure-electronic spin-orbit coupling parameters $\zeta$, $\mu$, and $\rho$, the spin gyro-magnetic ratios $g_s^{\mathrm{g}}$ and $g_s^{\mathrm{e}}$, and the orbital gyro-magnetic ratio $g_L$. Below we show how the values of these parameters can be found from existing spectral measurements under two different interpretations, the weak and strong DJT distortion. 

First, we find the relevant model parameters for the case of weak DJT distortion. Using the energy level assignments from the literature~\cite{Pappalardo1961OpticalI,Manson1971One-phononMgO,Moncorge1988ESA}, combined with our value for the ground-state spin g-factor $g_s^\mathrm{g}$, one could find values for the first six parameters of the model~\cite{Kaufmann1973DJT}, as detailed in Table~\ref{tab:params}. The value of $g_s^{\mathrm{e}}$ can be found from the value of $\zeta$ as shown in Appendix~\ref{app:T1_gfactor}, and $g_L=-\nicefrac{1}{2}$ can be calculated from the orbital structure, as shown in Appendix~\ref{app:gL}. 
\begin{table*}[tbh]
    \caption{List of SO and DJT parameter values and their derivations from independently measured values assuming a \emph{weak} DJT distortion.}
\begin{ruledtabular}
    \centering
    \begin{tabular}{|c|c|c|c|c|}
       Parameter & Value & Relation to measured values & Measured values &  Refs. \\
       \hline
       \hline
        $g_s^\mathrm{g}$ &2.242 &EPR and magneto-spectroscopy & 2.242 & This work \\
        \hline
        $\hbar\omega_\mathrm{ph}$ & 6.15~THz & Optical spectroscopy & 6.15~THz & \cite{Manson1971One-phononMgO} \\
        \hline
        $\zeta$ & -7.73~THz & $\zeta=(g_0-g_s^\mathrm{g})E_{ge}/4g_0$ & $E_{ge}=258$~THz\footnote{`Center of mass' of the optical absorption line} & \cite{Walsh1961StrainESR,Pappalardo1961OpticalI} \\
        \hline
        $E_\mathrm{JT}$ & 0.26~
        THz & $E_\mathrm{JT}=-(2/3)\hbar\omega_\mathrm{ph}\ln{(-\Delta_{32}/\zeta)}$ & $\Delta_{32}=7.26$~THz\footnote{Splitting between the $T_{2u}$ and $T_{1u}$ lines} & \cite{Moncorge1988ESA} \\
        \hline
        $\mu$ & 4.92~THz & $\mu=(\Delta_{43}+\Delta_{21}+e^{-\kappa/2}\zeta/2-3K_1)/3e^{-\kappa/2}$ & 
        $\Delta_{43}=12.6$~THz\footnote{Splitting between the $A_{2u}$ and $T_{2u}$ lines} & \cite{Moncorge1988ESA} \\
        \hline
        $\rho$ & -5.74~THz & $\rho=-\Delta_{21}-(1-e^{-\kappa/2})\mu-K_2$ & 
        $\Delta_{21}=5.28$~THz\footnote{Splitting between the $T_{1u}$ the $E_u$ lines} & \cite{Pappalardo1961OpticalI} \\
    \end{tabular}
    \label{tab:params}   
       \end{ruledtabular} 
\end{table*}

The obtained values of $E_\mathrm{JT}$, $\hbar\omega_\mathrm{ph}$, $\zeta$, and $g_L$ then yield $\kappa=0.13$, $K_1=0.14$~THz, and $K_2=0.15$~THz, consistent with a weak DJT distortion.

For the second case, for which $\kappa\gg1$, the only relevant spectroscopic data is the splitting between the two narrow, low-energy lines, $\Delta_{21}$. In this limit, $K_1,K_2\rightarrow0$, and thus $\Delta_{32}\approx-2\zeta e^{-\kappa/2}$. That is, it is becomes very small and may be below the resolution limit. Similarly, $\Delta_{43}\approx-\Delta_{21}-\zeta e^{-\kappa/2}$ and becomes indiscernible from $-\Delta_{21}$. With these approximated values one can see that $\mu$ has to be 0, and $\rho$ has to be $-\Delta_{21}$. All the other parameters are independent of the optical spectrum and can thus take the same values as in Table~\ref{tab:params}. 

We thus see that the same general model, though with different $\mu$ and $\rho$ values, can fit the available spectroscopic data also if we assume that the DJT distortion is strong. 

It is therefore impossible to discern between the case of weak DJT distortion and that of strong DJT distortion given only this data. For doing that, the additional measurements performed in this work were necessary. \color{black}

\setcounter{equation}{0}
\setcounter{table}{0}
\section{ESEEM for \texorpdfstring{$S=1$}{S=1} and \texorpdfstring{$I=\nicefrac{5}{2}$}{I=5/2}}\label{app:ESEEM}
  The Hamiltonian for an electron-spin--nuclear-spin interaction involving electron spin-states Zeeman-split in a magnetic field along the $z$-direction is diagonal in the electron spin-projection basis. For a given spin-projection value, $m_s$, it can be written as~\cite{Mims1965ESEEM,Probst2020ESEEM}
\begin{equation} \label{eq:H_SI1}
H_{m_s}=m_s\hbar\omega_S-\hbar\omega_II_z+m_sA_{zz}I_z+m_sA_{zx}I_x,
\end{equation}
where the magnetic field vector together with the vector pointing from the electron spin to the nuclear spin define the $x-z$ plane.

Thus, when considering only the $m_s=-1$ lower-state and the $m_s=0$ upper-state, which is a good approximation for the 9~mK experiment temperature, we obtain
\begin{equation}\label{eq:H_I1}
H_{-1}=-\hbar\omega_s-(\hbar\omega_I+A_{zz})I_z-A_{zx}I_x, 
\end{equation}
and 
\begin{equation}\label{eq:H_I0}
H_0=-\hbar\omega_II_z. 
\end{equation}   
It is clear that, while in the lower electronic spin state manifold there is coupling between the nuclear spin states, no such coupling is present in the higher electronic spin state manifold, where $m_s=0$, and the separation between the states in that manifold is determined only by the nuclear Zeeman frequency $\omega_I$, even if $A_{zz}$ and $A_{zx}$ are non-zero.  

At the temperature of the experiment, while most of the population is in the ground electronic spin state, the nuclear spin state is still fully mixed. Therefore, transitions starting in different ground states would not interfere. Hence, the frequencies of interference fringes are determined only by the frequency differences in the excited state, and will therefore be harmonics of the nuclear Zeeman splitting.
This is in striking contrast to the more commonly-studied case of $m_s=\pm\nicefrac{1}{2}$~\cite{Mims1965ESEEM,Probst2020ESEEM}.

For a localized electronic spin with an isotropic g-factor $g_s^\mathrm{g}$ interacting with neighboring nuclear spins, the contact interaction is negligible, and the components of the electron-spin--nuclear-spin interaction tensor are given by the dipole-dipole interaction~\cite{Probst2020ESEEM}
\begin{equation}
A_{ij}=\frac{3\mu_0}{4\pi|r|^5}\mu_\mathrm{B}\mu_Ng_s^\mathrm{g}g_n(|r|^2\delta_{ij}-3r_ir_j), 
\end{equation}
where $\mu_0$ is the permeability of vacuum, $\mu_N$ and $g_n$ are the nuclear magneton and the nuclear g-factor, respectively, and $\vec{r}$ is the position vector of the nuclear spin with respect to the electron spin. For MgO, the nearest Mg neighbors to the Ni substitutional site are located in all 12 permutations and sign combinations of $\vec{r}=a(\nicefrac{1}{2},0,\nicefrac{1}{2})$, where $a=0.42$~nm is the lattice constant of MgO. The length of all these vectors is $|r|=a/\sqrt{2}$. Out of the 12 possibilities, four do not contain the $z$-component, and for them $A_{zx}=0$, such that they won't exhibit any coupling between the different nuclear spin states, and won't contribute to ESEEM. For the other eight,
\begin{equation}
\begin{tabular}{ll}
$A_{zz}=$&$-\frac{3\sqrt{2}\mu_0}{4\pi a^3}\mu_\mathrm{B}\mu_Ng_s^\mathrm{g}g_n,$\\
 $A_{zx}=$& $\pm3A_{zz}.$
\end{tabular}
\end{equation}
  The minus sign in $A_{zx}$ appears in the 4 cases in which the $z$ coordinate is negative with respect to the direction of the magnetic field.  
Using $g_n=-0.34$ for $^{25}$Mg~\cite{NucGfactors1989}, we obtain   $A_{zz}/h=306$~kHz and $A_{zx}/h=\pm918$~kHz,   where $h$ is Planck's constant. These values are comparable in absolute value to the nuclear Zeeman splitting at $B=141$~mT, $\omega_I/2\pi=-366$~kHz (here the minus sign is due to the negative nuclear g-factor), leading to a significant modulation depth. For the simple case of $I=\nicefrac{1}{2}$, the visibility (which is only due to the nuclear spin state mixing in the $m_s=-1$ state), is given by~\cite{Probst2020ESEEM}
\begin{equation}\label{eq:V}
V=P_{25}V_1,
\end{equation}
where
\begin{equation}\label{eq:As}
V_1=\frac{|A_{zx}|/2}{\sqrt{(\hbar\omega_I+A_{zz})^2+A_{zx}^2/4}}.
\end{equation}
For the calculated interaction elements, $V_1=0.99$.
$P_{25}$ is the probability of a magnesium atom within the region affecting the nickel ion to be $^{25}$Mg. For example, if one considers only nearest neighbours, $P_{25}=1-(1-p_{25})^{N_{nn}}=0.57$.  Here $p_{25}=0.1$ is the natural abundance of $^{25}$Mg and $N_{nn}=8$ is the number of relevant nearest neighbour sites. As not only nearest neighbors contribute, in practice this number may be effectively closer to 1. The measured visibility at short times (before the onset of nuclear spin dephasing) is 0.75, which is in line with the above analysis. 

As the nuclear spin of $^{25}$Mg is $\nicefrac{5}{2}$ and not $\nicefrac{1}{2}$, additional modulation frequencies, harmonics of $\omega_I$ up to the fifth harmonic, are possible. Indeed, this is seen in the measurement [Fig.~\ref{fig:echo}(b)]. In order to calculate the ratios between the magnitude of the oscillations in the different frequencies, we numerically diagonalize the full $I=\nicefrac{5}{2}$ Hamiltonian, Eq.~(\ref{eq:H_I1}), using the same parameter values of Eq.~(\ref{eq:As}), and calculate the relative transition amplitudes using $S_x$   (the $x$ component of the electronic spin)   as the transition operator~\cite{Mims1965ESEEM,Probst2020ESEEM}. Then, for every ground state we sum the amplitudes leading to all excited states and calculate the transition probability at each transition frequency for that ground state. Finally, we sum over the probabilities calculated in this way for all ground states. The result, including bandwidth limitations due to the pulse duration and cavity width, are presented by the yellow bars in Fig.~\ref{fig:echo}(b).

In order to account for decoherence effects, we construct the temporal dependence of the transition probability from the result of the full $I=\nicefrac{5}{2}$ model described above, $p(t)$, and decompose it into its average value, $p_{\text{mean}}$, and a purely oscillating component, $p_{\text{osc}}(t)$.
We then produce the following function, 
\begin{equation}
    p_{\text{echo}}(t)=g_1(t)[p_{\text{mean}}+g_2(t)p_{\text{osc}}(t)],
\end{equation}
where 
\begin{equation}
\begin{tabular}{lcl}
    $g_1(t)$&$=$&$Ae^{-\nicefrac{t}{T^{\text{(short)}}_2}}+Be^{-\nicefrac{t}{T^{\text{(long)}}_2}}$,\\
    $g_2(t)$&$=$&$P_{25}e^{-\nicefrac{t}{T^{*\text{(nuc)}}_2}}$.
\end{tabular}
\end{equation}
The result of fitting this function to the data is presented by the solid black line in Fig.~\ref{fig:echo}(a). Note that here we used the measured $|\omega_I/2\pi|=385$~MHz instead of the calculated value, and used $P_{25}$ as a fit parameter. Table~\ref{tab:fit_params} presents the fitted parameter values.
\begin{table}[tbh]
    \caption{List of fit parameters for the spin-echo trace presented in Fig.~\ref{fig:echo}(a).}
\begin{ruledtabular}
    \centering
    \begin{tabular}{|c|c|}
       Parameter & Value \\
       \hline
       \hline
        $A$ & $0.917\pm0.004$ \\
        \hline
        $B$ & $0.0825\pm0.0005$ \\
        \hline
        $P_{25}$ & $0.81\pm0.005$ \\
        \hline
        $T_2^{\text{(short)}}$ & $4.50\pm0.03$~$\mu$s\\
        \hline
        $T_2^{\text{(long)}}$ & $109\pm2$~$\mu$s\\
        \hline
        $T_2^{*\text{(nuc)}}$ & $52\pm2$~$\mu$s\\ 
    \end{tabular}
    \label{tab:fit_params}   
       \end{ruledtabular} 
\end{table}

\setcounter{equation}{0}
\setcounter{table}{0}
\section{Spectral diffusion dephasing for \texorpdfstring{$S=1$}{S=1}}\label{app:SpectDiff}
Spectral diffusion of the probed spins can be caused by flip-flop processes within the surrounding spin bath, which stochastically change the magnetic field environment of the probed spins. The resulting dephasing rate is proportional to the number of spin pairs that can flip-flop. As the flip-flop process conserves energy, one should count only the spin pairs in which the two spins have the same energy splitting. This number depends on the spin populations and hence depends on temperature, making the spectral diffusion dephasing rate temperature-dependent. 

In order to find the temperature dependence, we first calculate the number of spin pairs at a given temperature, for a certain strain detuning of the $m_s=0$ level, $\delta$. We note that in a S=1 system there are three possible flip-flop processes: 
\begin{equation}
\begin{tabular}{lcr}
$|1,0\rangle$&$\leftrightarrow$&$|0,1\rangle,$\\ $|-1,0\rangle$&$\leftrightarrow$&$|0,-1\rangle,$\\  $|1,-1\rangle$&$\leftrightarrow$&$|0,0\rangle,$ 
\end{tabular}
\end{equation}
where $|m_{s,1},m_{s,2}\rangle$ represents a state of a spin pair in levels $m_{s,1}$ and $m_{s,2}$. The numbers of spin-pairs that conserve energy for these processes are, respectively,
\begin{equation}
\begin{tabular}{c}
$n_1(\delta)n_0(\delta),$\\
$n_{-1}(\delta)n_0(\delta),$\\
$\tfrac{1}{2}[n_{-1}(\delta)n_1(-\delta)+n_0(\delta)n_0(-\delta)],$\\
\end{tabular}
\end{equation}
where $n_{m_s}(\delta)$ is the number of spins (per unit detuning) populating the level $m_s$ for a system with $m_s=0$ detuning of $\delta$. 
At a given temperature, the populations of the three levels are 
\begin{equation}
\begin{tabular}{lcr}
$n_{-1}(\delta)$&$\propto$&$1/Z(\delta),$\\
$n_{0}(\delta)$&$\propto$&$e^{-\beta(\varepsilon_B-\delta)}/Z(\delta),$\\
$n_{1}(\delta)$&$\propto$&$ e^{-2\beta\varepsilon_B}/Z(\delta),$\\
\end{tabular}
\end{equation}
where here $\beta=1/k_BT$ is the inverse temperature in units of energy ($k_B$ is the Boltzmann constant) and $\varepsilon_B=g_s^\mathrm{g}\mu_\mathrm{B}B$ is the Zeeman energy splitting. $Z(\delta)=1+e^{-\beta(\varepsilon_B-\delta)}+e^{-2\beta\varepsilon_B}$ is the partition function. Note that here we neglected any strain shift of the $m_s=1$ level. 

The total spectral diffusion dephasing rate is then given by summing the number of pairs for the three processes for a given $\delta$, multiplying by the detuning distribution function $P(\delta)=\tfrac{1}{\sqrt{2\pi}\sigma}e^{-\delta^2/2\sigma^2}$ ($\sigma$ being the width of the distribution), and integrating over $\delta$. Adding also a temperature independent dephasing rate, representing the direct flip-flop and instantaneous diffusion processes, which are effectively constant at the (low) temperatures where they are significant, this yields 
\begin{widetext}
\begin{equation}
\frac{1}{T_2}=\frac{1}{T_{2,LT}}+\frac{3}{T_{2,SD,\mathrm{sat}}}\int_{-\infty}^{\infty}d\delta P(\delta)\left(\frac{e^{-\beta(\varepsilon_B-\delta)}+e^{-\beta(3\varepsilon_b-\delta)}}{Z^2(\delta)}+\frac{e^{-2\beta\varepsilon_B}}{Z(\delta)Z(-\delta)}\right),
\end{equation}
\end{widetext}
where $T_{2,LT}$ is the low-temperature dephasing time and $T_{2,SD,\mathrm{sat}}$ is the spectral diffusion dephasing time at saturation (note that for low temperatures the integral tends to 0, while for high temperatures it tends to $\nicefrac{1}{3}$), and we used the fact that $P(\delta)$ is even.

For the fit used in Fig.~\ref{fig:decays}, we used the measured values of $\sigma=170$~MHz and $\varepsilon_B=4386$~MHz ($B=141$~mT), and fitted the values of $T_{2,LT}$ and $T_{2,SD,\mathrm{sat}}$ to the data. We obtain $T_{2,LT}=85\pm10$~$\mu$s and $T_{2,SD,\mathrm{sat}}=2.65\pm0.2$~$\mu$s.

\setcounter{equation}{0}
\section{Spin g-factor in the \texorpdfstring{$^3$T$_{2g}$$(T_{1u})$}{3T2g(T1u)} states}\label{app:T1_gfactor}
The SO interaction may alter the value of the spin g-factor of a state by mixing it with other states of the same SO representation~\cite{Misetich1964gfactor}. For the $T_{1u}$ SO states of the $^3$T$_{2g}$ manifold, the closest such states are the $T_{1u}$ SO states of the $^3$T$_{1g}$ manifold. We thus assume a wave-function of the form,
\begin{equation} 
|\psi_j\rangle=N_j\left(|T_{1u,j}\rangle_{^3\mathrm{T}_{2g}}+\sum_{k=-1}^1\gamma_{k,j}|T_{1u,k}\rangle_{^3\mathrm{T}_{1g}}\right),
\end{equation} 
where $\gamma_{k,j}$ is the `mixing fraction' of the state $|T_{1u,k}\rangle_{^3\mathrm{T}_{1g}}$ in $\psi_j$, and $N_j=(1+\sum_{k=-1}^1\gamma_{k,j}^2)^{-1/2}$ is a normalization constant. Using second-order perturbation theory, the mixing fractions are given by,
\begin{equation}
\gamma_{k,j}=\frac{_{^3\mathrm{T}_{2g}}\langle T_{1u,k}|H_{SO}|T_{1u,j}\rangle_{^3\mathrm{T}_{1g}}}{E_{^3\mathrm{T}_{1g}}^{T_{1u}}-E_{^3\mathrm{T}_{2g}}^{T_{1u}}},
\end{equation}
where $H_{SO}$ is the SO Hamiltonian, and $E_{^3\mathrm{T}_{1g}}^{T_{1u}}$ ($E_{^3\mathrm{T}_{2g}}^{T_{1u}}$) is the energy of the $T_{1u}$ states in the $^3\mathrm{T}_{1g}$ ($^3\mathrm{T}_{2g}$) manifold. The effective g-factor is then given by 
\begin{equation}
g_{T_{1u}}=\tfrac{1}{\hbar}\langle \psi_1|L_z+g_0S_z|\psi_1\rangle.
\end{equation}
Using the known structure of the $T_{1u}$ wave-functions~\cite{Griffith1961} and assuming $\gamma_{k,j}\ll1$, one obtains,
\begin{equation}\label{eq:gT1u}
g_{T_{1u}}\approx\tfrac{1}{2}\left[g_L+g_0(1-2\gamma_{1,-1})\right],
\end{equation}
where $g_L=\langle\mathrm{T}_{2g,1}|L|\mathrm{T}_{2g,1}\rangle$ is calculated in Appendix~\ref{app:gL} below. Note that this calculation does not include DJT distortion. Its inclusion amounts to replacing $g_L$ with $\tilde{g}_L$. By table \ref{tab:weak_sr} we identify 
\begin{equation}
g_s^{\mathrm{e}}=g_0(1-2\gamma_{1,-1}).
\end{equation}
Using $E_{^3\mathrm{T}_{1g}}^{T_{1u}}-E_{^3\mathrm{T}_{2g}}^{T_{1u}}=160$~THz~\cite{Moncorge1988ESA}, $_{^3\mathrm{T}_{2g}}\langle T_{1u,k}|H_{SO}|T_{1u,j}\rangle_{^3\mathrm{T}_{1g}}=-\frac{\sqrt{3}}{2}\zeta$~\cite{Liher1959CFTmatrices,Griffith1961}, and the value of $\zeta$ extracted from the g-factor of the ground-state (Appendix~\ref{app:DJTmodel}), $\zeta=-7.73$~THz, we obtain $\gamma_{1,-1}=0.04$, yielding $g_s^{\mathrm{e}}=1.84$.

\setcounter{equation}{0}
\section{Orbital g-factor in the \texorpdfstring{$^3$T$_{2g}$}{3T2g} manifold}\label{app:gL}
In the $d^8$ configuration, it is much easier to use the hole notation, as then one has to consider only the two empty electron orbitals instead of the eight full ones. We therefore first find the orbital angular momentum of the two-electron orbitals from which electrons are missing, and then take the negative of the result, as we are interested in the orbital angular momentum of a full shell (which is 0) \emph{minus} those two electrons. 

All the relevant two-electron orbitals are composed mostly of products of two $d$ ($l=2$) single-electron states. Therefore, they are mostly composed of doubly-quadratic functions of the Cartesian coordinates.

For spin-triplet states, which are exchange-symmetric, the orbitals also have to be exchange-anti-symmetric, as the total electronic wave-function must be exchange-anti-symmetric.

Out of all the possible T$_{2g}$ two-electron orbitals, we focus here on those in which one electron is of the $e_g$ single-electron orbital and the other is of the  $t_{2g}$ single-electron orbital [see Fig.~\ref{fig:levels}(b)]. 

Therefore, we have to use a set of spanning functions of the T$_{2g}$ representation of the $O_h$ point group, which are doubly-quadratic in $x$, $y$, and $z$, are composed of products of quadratic basis functions of the E$_g$ and T$_{2g}$ representations, and are exchange-anti-symmetric. 

These constraints leave only one possible choice (up to internal unitary transformations). Choosing the main axis to be $z$ and all the functions to be eigen-functions of the $z$ component of the angular momentum, we are left with~\cite{Griffith1961}
\begin{equation}
\begin{tabular}{lll}
$|$T$_{2g,1}\rangle_{\mathrm{e}}$&$=$&$-\tfrac{a}{\sqrt{2}}\{x_1z_1(3y_2^2-r_2^2)+iy_1z_1(3x_2^2-r_2^2)\}_\mathrm{a.s.},$\\
$|$T$_{2g,0}\rangle_{\mathrm{e}}$&$=$&$ia\{x_1y_1(3z_2^2-r_2^2)\}_\mathrm{a.s.},$\\
$|$T$_{2g,-1}\rangle_{\mathrm{e}}$&$=$&$\tfrac{a}{\sqrt{2}}\{x_1z_1(3y_2^2-r_2^2)-iy_1z_1(3x_2^2-r_2^2)\}_\mathrm{a.s.},$
\end{tabular}
\end{equation}
where the subscript $\mathrm{e}$ indicates that these are electronic orbitals, $\{\ \}_\mathrm{a.s.}$  stands for exchange-anti-symmetrization, $r_{1(2)}^2=x_{1(2)}^2+y_{1(2)}^2+z_{1(2)}^2$, and $a=5/(4\pi\sqrt{2})$  is a normalization constant.

By inverting the definitions of the spherical harmonics in Cartesian coordinates, it can be shown~\cite{Griffith1961} that these functions can be represented as the following combinations of products of single-electron $d$ orbitals,
\begin{equation}
\begin{tabular}{lll}
$|$T$_{2g,1}\rangle_{\mathrm{e}}$&$=$&$\{\tfrac{\sqrt{3}}{\sqrt{8}}(|1,2\rangle+|1,\mathrm{-}2\rangle)-\tfrac{1}{2}|\mathrm{-}1,0\rangle\}_\mathrm{a.s.},$\\
$|$T$_{2g,0}\rangle_{\mathrm{e}}$&$=$&$\tfrac{1}{\sqrt{2}}\{|2,0\rangle-|\mathrm{-}2,0\rangle\}_\mathrm{a.s.},$\\
$|$T$_{2g,-1}\rangle_{\mathrm{e}}$&$=$&$-\{\tfrac{\sqrt{3}}{\sqrt{8}}(|\mathrm{-}1,2\rangle+|\mathrm{-}1,\mathrm{-}2\rangle)-\tfrac{1}{2}|1,0\rangle\}_\mathrm{a.s.},$
\end{tabular}
\end{equation}
where the two numbers in the kets stand for the eigenvalues of the $z$ component of the angular momentum of the single-electron $d$ orbitals of the two electrons. 

With the wave-functions cast in this form, it is straightforward to calculate their two-electron angular momentum matrix elements. One obtains,
\begin{eqnarray}
\langle L_x\rangle=\frac{\hbar}{2\sqrt{2}}\left(\begin{array}{ccc}
0 &1 &0 \\
1 &0 &1 \\
0 &1 &0 
\end{array}\right),\\ 
\langle L_y\rangle=\frac{\hbar}{2\sqrt{2}}\left(\begin{array}{ccc}
0 &-i &0 \\
i &0 &-i \\
0 &i &0 
\end{array}\right),\\ 
\langle L_z\rangle=\frac{\hbar}{2}\left(\begin{array}{ccc}
1 &0 &0 \\
0 &0 &0 \\
0 &0 &-1 
\end{array}\right).
\end{eqnarray}

It is clear that these are spin-1 matrices, as expected, just multiplied by a common factor of $\nicefrac{1}{2}$. Thus, one can treat the T$_{2g}$ two-electron orbitals as an effective $l=1$ system, with an effective orbital g-factor of $\nicefrac{1}{2}$. 

Now, recall that we are interested in the orbital angular momentum of a full $d$ shell \emph{missing} two electrons in the above states. We should therefore take the \emph{negative} of the above result. That is, 
\begin{equation}
g_L=-\nicefrac{1}{2}.
\end{equation}


\bibliography{NiMgO}

\end{document}